\documentclass[journal=jacsat,manuscript=article]{achemso}

\usepackage[version=3]{mhchem} 

\usepackage{graphicx}
\usepackage{color}
\usepackage{dcolumn} 
\usepackage{bm}      
\usepackage{amssymb}
\usepackage{amsfonts}
\usepackage{svg}
\usepackage{comment}
\usepackage{xr}
\usepackage[bookmarks=false,colorlinks,citecolor=red]{hyperref}
\usepackage{float}
\usepackage{caption}
\usepackage{amsmath}
\usepackage{mhchem}
\usepackage{subcaption}
\usepackage[colorlinks,citecolor=red]{hyperref}
\usepackage{multicol}
\usepackage{multirow}
\usepackage{algorithm}
\usepackage{algpseudocode}
\usepackage[normalem]{ulem}
\usepackage[symbol]{footmisc}
\SectionNumbersOn

\newcommand{\COMMENTED}[1]{}

\DeclareMathOperator*{\argmax}{arg\,max}
\DeclareMathOperator*{\argmin}{arg\,min}

\makeatletter
\newcommand*{\addFileDependency}[1]{
  \typeout{(#1)}
  \@addtofilelist{#1}
  \IfFileExists{#1}{}{\typeout{No file #1.}}
}
\makeatother

\newcommand*{\myexternaldocument}[1]{%
    \externaldocument{#1}%
    \addFileDependency{#1.tex}%
    \addFileDependency{#1.aux}%
}

\myexternaldocument{supporting}

\author{Gopal R. Iyer}
\affiliation{Department of Chemistry, Brown University, Providence, Rhode Island 02912, USA}
\author{Noah Whelpley}
\affiliation{Department of Chemistry, Brown University, Providence, Rhode Island 02912, USA}
\author{Juha Tiihonen}
\affiliation{Department of Physics, Nanoscience Center, University of Jyv\"{a}skyl\"{a}, Jyv\"{a}skyl\"{a}, Finland}
\author{Paul R. C. Kent}
\affiliation{Computational Sciences and Engineering Division, Oak Ridge National Laboratory, Oak Ridge, Tennessee 37831, USA}
\author{Jaron T. Krogel}
\affiliation{Materials Science and Technology Division, Oak Ridge National Laboratory, Oak Ridge, Tennessee 37831, USA}
\email{krogeljt@ornl.gov}
\author{Brenda M. Rubenstein}
\affiliation{Department of Chemistry, Brown University, Providence, Rhode Island 02912, USA}
\email{brenda_rubenstein@brown.edu}

\title[]
  {Force-free identification of minimum-energy pathways and transition states for stochastic electronic structure theories}

\begin{document}


%




\begin{abstract}
 The accurate mapping of potential energy surfaces (PESs) is crucial to our understanding of the numerous physical and chemical
 processes mediated by atomic rearrangements - such as conformational changes and chemical reactions - and the thermodynamic and
 kinetic feasibility of these processes. Stochastic electronic structure theories, e.g., Quantum Monte Carlo (QMC) methods, enable
 highly accurate total energy calculations which in principle can be used to construct the PES. However, their stochastic nature poses a
 challenge to the computation and use of forces and Hessians, which are typically required in algorithms for minimum-energy pathway (MEP)
 and transition state (TS) identification, such as the nudged-elastic band (NEB) algorithm and its climbing image formulation. Here,
 we present strategies that utilize the surrogate Hessian line-search method - previously developed for QMC structural optimization
 - to efficiently identify MEP and TS structures without requiring force calculations at the level of the stochastic electronic
 structure theory. By modifying the surrogate Hessian algorithm to operate in path-orthogonal subspaces and on saddle points, we
 show that it is possible to identify MEPs and TSs using a force-free QMC approach. We demonstrate these
 strategies via two examples, the inversion of the ammonia (NH$_3$) molecule and the nucleophilic substitution (S$_\textrm{N}$2)
 reaction F$^-+$ CH$_3$F $\rightarrow$ FCH$_3$ + F$^-$. We validate our results using Density Functional Theory (DFT)- and coupled
 cluster-based NEB calculations. We then introduce a hybrid DFT-QMC approach to compute thermodynamic and kinetic quantities - free
 energy differences, rate constants, and equilibrium constants - that incorporates stochastically-optimized structures and their
 energies, and show that this scheme improves upon DFT accuracy. Our methods generalize straightforwardly to
 other systems and other high-accuracy theories that similarly face challenges computing energy gradients,
 paving the way for highly accurate PES mapping, transition state determination, and thermodynamic and kinetic
 calculations, at significantly reduced computational expense.
\end{abstract}

\section{Introduction}
One of the crowning achievements of modern quantum chemistry is the ability to compute potential energy surfaces (PESs) entirely from first principles.~\cite{cramer_computational_chemistry, pes_1} With knowledge of these surfaces, researchers can predict the reactivity, dynamics, and kinetics of a wide range of technologically important and scientifically interesting atomistic systems.~\cite{pes_2, high_dimensional_pes} For instance, the stable structures of the reactants and products of a catalytic reaction can be determined by identifying the minima on its PES using a variety of geometry optimization algorithms.~\cite{geometry_optimization_review, gpmin_geometry_optimization, fire_geometry_optimization} Reaction mechanisms can, moreover, 
be predicted by determining the minimum-energy pathway (MEP) that connects the reactant and product states of a reactive system using, for instance, chain-of-states methods\cite{chain_of_states_1, chain_of_states_2} or the nudged elastic band (NEB) method\cite{neb_original_paper} and its variations\cite{neb_2, neb_3}. These methods can often be straightforwardly extended - e.g., using the climbing-image nudged elastic band (CI-NEB) method\cite{ci_neb} - to identify the (usually) first-order saddle point of the PES along the MEP, which corresponds to the transition state. Reaction mechanisms predicted in this fashion provide invaluable atomistic information about reaction pathways, granting researchers key insights into how reactions can be redesigned and potentially controlled to minimize energy, material, and other resource requirements. Indeed, the ability to computationally predict reaction mechanisms has become an indispensable tool in the fields of combustion analysis,\cite{combustion} catalysis,\cite{heterogeneous_catalysis_1, heterogeneous_catalysis_2} and chemical reactor design.~\cite{martinez_nanoreactor}

Nonetheless, while virtually all quantum chemical methods can predict the energy of a chemical system given the geometries of the species it contains, it can be challenging for some methods to do the inverse, i.e., to very accurately predict equilibrium (reactant/product) and transition states, and reaction pathways given the energies of structures sampled in some local region of the PES. This is because being able to efficiently determine equilibrium structures, transition states, and MEPs requires knowledge of energy derivatives, including forces (first derivatives) or Hessian matrices (second derivatives). These quantities then guide the dynamics of initial structures to the true equilibrium or saddle point structures that are consistent with the given level of \textit{ab initio} theory. 
Computing such derivatives in methods like Density Functional Theory (DFT) can be done in a relatively straightforward fashion using the Hellman-Feynman Theorem,\cite{hellmann_feynman} making DFT one of the most popular methods for determining molecular structures. However, computing these derivatives in more expensive many-body theories, including Coupled Cluster Theory (particularly with the inclusion of single, double, and perturbative triple excitations, CCSD(T)),\cite{Salter_JCP_1989,Stanton_Gauss_2000}
the Random Phase Approximation (RPA),\cite{burow_analytical_2014,Kresse_RPA_Forces} or Quantum Monte Carlo (QMC)\cite{qmc_force_1,qmc_force_2,qmc_force_3,Filippi_PRB_2000,Assaraf_Cafferel_2003,Rios_PRE_2019,Nakano_Sorella_2022,moroni_practical_2014,qmc_force_4,chen_structural_2022} methods can be either approximate or highly costly. In the case of stochastic many-body theories such as QMC, this also requires formal treatment of the energy derivatives in the presence of inherent statistical errors in the energies.~\cite{qmc_force_1} As a result, instead of determining these geometries or paths using highly accurate many-body methods, many resort to using potentially less accurate theories like DFT to determine structures and then higher-accuracy theories to determine their corresponding energies.~\cite{finite_size_error_cancellation} Although DFT geometries can be sufficiently accurate in the many cases in which electron correlation does not play a critical role in dictating geometries, computing properties using many-body theories based on single-reference geometries can lead to a fundamental mismatch in theories that can result in inaccurate or inconsistent predictions. For instance, as the authors have previously shown, lower-accuracy DFT geometries of 2D materials that can be readily stretched/compressed along particular axes can yield misleading predictions of magnetic and other electronic properties when studied using more accurate quantum chemical methods.~\cite{CrI3_dmc, GeSe_dmc} It would thus be ideal to use the same high-accuracy method to both predict structures and their energies and other properties.

Recently, a variety of hybrid approaches for circumventing and reducing the cost of directly computing energy gradients using relatively expensive high-accuracy methods have been proposed. Several approaches, particularly machine learning (ML)-based methods for mapping PESs and computing QMC forces, have been demonstrated successfully.~\cite{qmc_ml_energies, qmc_force_5, transfer_learning_potential,Archibald_JCP_2018} While such ML methods open up exciting new avenues for the scalable modeling of large molecules and complex chemical reactions,~\cite{Archibald_JCP_2018} they are typically highly reliant on the generation of large training sets. This can limit their practicality for systems for which it is difficult to assemble sufficient data, curtailing their generalizability beyond the systems on which they have been trained. Furthermore, the accuracy of ML methods can be significantly improved if they are trained with both energies \emph{and} forces,\cite{qmc_force_5,ceperley2023training} which may be challenging to accomplish if forces are not readily provided by the methods used to furnish training data. 

An alternative set of approaches that do not necessitate large training sets or forces are so-called surrogate approaches, which use information easier to obtain from less accurate, but also less expensive, theories to guide more accurate, more expensive theories. Some of the first ideas along these lines were proposed in Ref.~\citenum{Grossman_PRL}. More recently,  Ref.~\citenum{original_paper} introduced the surrogate Hessian line-search method, which locates energy minima to within some tunable structural error tolerance via a series of one-dimensional line searches that account for and leverage statistical errors in QMC energy calculations. This method was then successfully applied to the structural optimization of both molecular\cite{original_paper} and material systems.~\cite{CrI3_dmc, GeSe_dmc}

Building upon the success of this approach for identifying equilibrium geometries, in this work, we develop new formalisms and algorithms for generalizing this approach to the determination of transition state geometries and other structures along the minimum energy pathways of chemical processes. Key to generalizing these techniques is the realization that reaction pathway identification tasks can be viewed as iterative energy optimization steps. Transition states (saddle points) can be identified by minimizing the energy along all directions except one that has negative curvature, along which the energy is maximized instead. MEPs can be identified by minimizing the energy along all directions in the subspace orthogonal to the tangent to the path at each structure along some guess discretized path. We note here that the idea of reformulating MEP identification as a series of energy optimization problems instead of force-based dynamical problems has been suggested before, notably in Ref.~\citenum{optimization_string_method}. Additionally, Saccani \textit{et al.} (Ref. \citenum{Saccani_JCP_2013}) demonstrated the application of the NEB algorithm to QMC-based MEP identification using correlated sampling.~\cite{Filippi_PRB_2000} However, to the best of our knowledge, ours is the first approach that generalizes MEP identification to stochastic electronic structure methods by utilizing Hessian information from lower-level surrogate theories, eliminating the need for computing derivatives at the QMC level. We demonstrate the accuracy and utility of this formalism by determining the transition states and minimum energy pathways of two paradigmatic chemical processes: (i) ammonia (NH$_3$) inversion, which involves two structural degrees of freedom, and (ii) and an S$_\textrm{N}$2 nucleophilic substitution reaction, F$^-$ + CH$_3$F $\rightarrow$ FCH$_3$ + F$^-$,\cite{reactions} that involves four degrees of freedom, using Diffusion Quantum Monte Carlo (DMC) informed by DFT Hessian computations. We compare the minimum energy pathways we compute using DMC with those obtained using DFT- and Coupled Cluster-based NEB calculations, demonstrating that our surrogate techniques can provide at least post-DFT quality pathways at only a small factor of $M\times (N-1)$ greater cost than traditional single-point DMC energy calculations, where $M$ is the number of points along which a discrete one-dimensional line-search grid is generated and $N$ is the number of structural parameters used to parameterize the system. We subsequently show that we can use a combined DFT-QMC approach to more accurately estimate thermodynamic quantities, such as free energies of reactions, and kinetic quantities, such as rate constants, which can often be more directly compared to experiments. Overall, this work provides clear avenues for determining reaction pathways and transition states with chemical (or greater) accuracy at substantially lower computational cost than direct approaches, but without the need for the large training datasets that statistical learning-based methods require. We believe that these techniques will provide new tools for the study of reactions and processes that involve significant amounts of electron correlation and the design of novel homogeneous and heterogeneous catalysts that can accelerate these reactions.

This manuscript is organized as follows. In Section \ref{sec:methods}, we introduce the theoretical foundations of the surrogate Hessian line-search method and then show how this formalism can be generalized to transition states (Section \ref{sec:eqm_and_saddle}) and minimum energy pathways (Section \ref{sec:subspace_optimization}). We also describe how thermodynamic corrections to obtain free energies and equilibrium constants can be performed using the hybrid DFT-QMC approach (Section \ref{sec:hybrid_dft_qmc}). We then present our results and benchmarks on our two illustrative chemical processes in Section \ref{sec:results_and_discussion}. Lastly, in Section \ref{sec:conclusions}, we conclude with the impacts and potential further generalizations of this method.

\section{Methods}
\label{sec:methods}
\subsection{Theoretical Background}
\label{sec:theoretical_background}
\subsubsection{Surrogate Hessian Line-Search Method for Equilibrium and Transition States}
\label{sec:eqm_and_saddle}
The original surrogate Hessian line-search method for identifying equilibrium structures (energy minima) has previously been introduced in Ref.~\citenum{original_paper}. For brevity, we only summarize the method here while noting the specific modifications necessary to extend the method to transition state identification.


In the surrogate Hessian line-search method, a surrogate method (such as DFT) that can more rapidly compute energy gradients is used to calculate Hessians that can then be used to determine the directions along which line searches for minima should be performed using energies from a higher-accuracy and costlier theory. This method relies on the assumption that the local curvature of the target, higher-accuracy PES near a critical (0-derivative) point - energy minimum or saddle point - is approximated reasonably well by the PES calculated using a surrogate theory. This allows one to compute (a) the optimal directions along which to perform the line-search, and (b) the maximum tolerable statistical noise (if a stochastic method is employed) at each point on the line-search grid in order to be able to determine an optimized geometry within some structural tolerance. This information is then used to perform a series of one-dimensional line-searches to locate the critical point within some confidence interval.

While the surrogate and higher accuracy approaches that can be employed are highly generalizable, throughout the rest of this manuscript, we focus on the specific choices of Density Functional Theory (DFT) as our surrogate method and DMC\cite{foulkes_qmc, ceperley_qmc} as our higher accuracy method. This choice of methods is natural since DFT calculations are often used as inputs into DMC calculations. In this context, the key steps of the surrogate Hessian line-search method can be summarized as follows:

\begin{enumerate}
    \item \textbf{Initialization and parameterization}: To provide an initial seed structure to the equilibrium or saddle point line-search, we use an optimized equilibrium or transition state geometry obtained using the surrogate theory (DFT). This may be the result of a typical structural relaxation or CI-NEB calculation, respectively. In systems where some degrees of freedom are correlated due to structural symmetries, it is often desirable to define a few structural parameters that capture these symmetries and only describe the relevant regions of configuration space instead of operating in the full space of $3N$ Cartesian coordinates (where $N$ is the number of atoms). We achieve this by defining a mapping function $\mathrm{F^{I}}:\mathbb{R}^{3N}\rightarrow \mathbb{R}^d$ (given $d$ structural parameters). It is defined to ensure that all structures that share these symmetries map to the same representation in $\mathbb{R}^d$, while its inverse ($(\mathrm{F^{I}})^{-1}$) maps a structure in $\mathbb{R}^d$ to a single structure in $\mathbb{R}^{3N}$. When working with solids, it is also reasonable to include the dimensions of the unit cell as additional structural parameters to allow for relaxations of the overall lattice. The structural parameterizations of the NH$_3$ inversion and S$_\textrm{N}$2 reaction systems studied in this manuscript are described in Section \ref{sec:testcases}.
    \item \textbf{Hessian evaluation}: Given a reduced structural representation of the initial state, we then compute the Hessian matrix of the PES in the vicinity of that structure using the surrogate theory. Here, since we only define a small number of degrees of freedom, we evaluate the Hessian using a finite-difference scheme. For more complex systems, it may be beneficial to evaluate it in generalized coordinates and then map it to the reduced space. Notably, when dealing with equilibrium geometries, we expect all the Hessian eigenvalues to be positive, while we expect $n$ eigenvalues to be negative when dealing with $n^\textrm{th}$-order saddle points. This distinction will become important when defining the objective function of the line-search. Furthermore, the Hessian eigenvectors provide near-optimal directions along which to perform the line-search using the stochastic theory.
    \item \textbf{Tolerance optimization and line-search}: The Hessian eigenvectors now provide us with directions along which we can perform a series of one-dimensional line-search operations to determine the desired equilibrium or saddle point. This requires evaluating energies using the stochastic theory at a few points along each line-search direction. To determine these points, it is necessary to identify the optimal (minimal) degree of stochastic energy sampling required to be able to specify the structural parameters with a certain pre-specified level of precision ($\delta P$). For each line-search direction, this is equivalent to evaluating the maximum point-wise energy noise tolerated along each line-search direction, $\sigma^\textrm{max}$, and the corresponding structural window along which the energy grid is sampled, $L^\textrm{max}$, both of which are functions of $\delta P$. Finally, along each line-search direction $x$, the objective of the line-search for determining the equilibrium geometry can be expressed simply as an energy minimization criterion
    \begin{equation}
        x_\textrm{min}^* = \argmin E(x).
        \label{eq:objective}
    \end{equation}
    The modification to this objective for locating a saddle point is then relatively straightforward. If a certain line-search direction corresponds to a negative eigenvalue, instead of minimizing the energy along that direction, we \textit{maximize} it. This satisfies the definition of a saddle point as lying at a local maximum on the PES along some (small number of) directions while lying at a local minimum along all others. The line-search objective is then modified as
    \begin{equation}
        x_\textrm{saddle}^* = \begin{cases} 
      \argmin E(x) & \text{if } \lambda_x > 0 \\
      \argmax E(x) & \text{if } \lambda_x < 0
      \label{eq:modified_objective}
   \end{cases},
    \end{equation}
    where $\lambda_x$ is the eigenvalue corresponding to the Hessian eigenvector $x$.
\end{enumerate}

We direct the reader to Ref.~\citenum{original_paper} for further details regarding the surrogate Hessian line-search method.

\subsubsection{Path-Orthogonal Subspace Optimization}
\label{sec:subspace_optimization}
We now describe how the surrogate Hessian line-search method can be adapted to identify MEPs. We begin by defining an MEP according to Fukui's formulation of the intrinsic reaction coordinate (see Ref.~\citenum{fukui_irc}) as a one-dimensional sub-manifold in configuration space that satisfies the following conditions:
(1) connects two equilibrium states;
(2) passes through a saddle point (assumed to be first-order for simplicity); and
(3) consists exclusively of points that lie in local energy minima in all directions except along the tangent to the path at those points.

\begin{figure}[ht]
    \centering
    \begin{subfigure}[b]{0.49\textwidth}{\includegraphics[width=\textwidth]{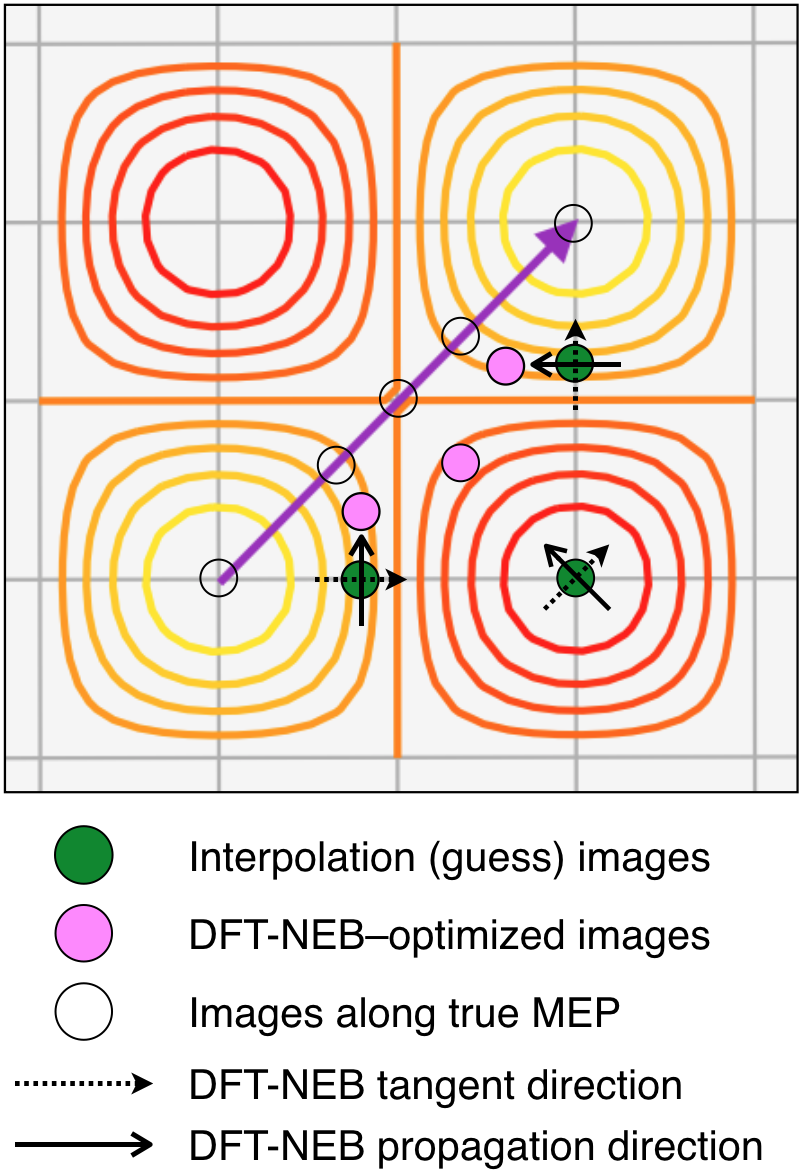}}
    \caption{}
    \label{fig:algorithm_left}
    \end{subfigure}
    \hfill
    \begin{subfigure}[b]{0.49\textwidth}{\includegraphics[width=\textwidth]{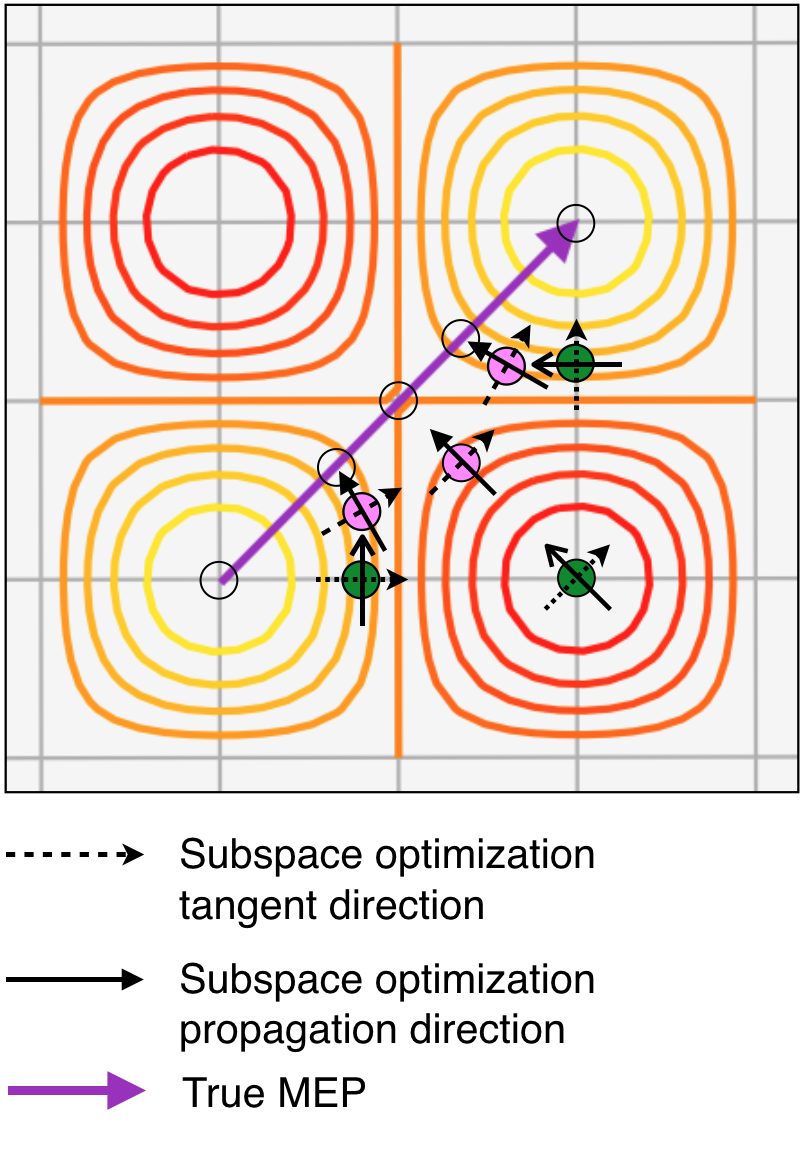}}
    \caption{}
    \label{fig:algorithm_right}
    \end{subfigure}
    \caption{Schematic illustration of the (a) NEB and (b) surrogate Hessian subspace optimization algorithms on a contour map of the true PES. Yellow and red contours indicate local energy minima and maxima, respectively.}
    \label{fig:algorithm}
\end{figure}

MEPs are typically identified using chain-of-states methods such as NEB (Fig. \ref{fig:algorithm_left}).~\cite{neb_original_paper} The NEB algorithm begins with the construction of an initial guess pathway, usually a linear interpolation of $M$ points between two equilibrium states (satisfying condition 1). For each of these pathway structures (images), tangents are calculated along the path. Then, two types of forces acting on (each atom of) each image are computed: (1) the ``true'' force, i.e., the negative gradient of the potential energy calculated according to the prescribed level of theory, and (2) an artificial ``spring'' force calculated as a function of the excess distance between adjacent images compared to some equilibrium distance, usually set to be $\mathcal{D}/(M+1)$, where $\mathcal{D}$ is the Euclidean distance between the initial and final states. (Note that the result of an NEB calculation ought to be agnostic to the spring constant, provided it is set to a reasonable value, typically $\sim 0.1$ eV/\AA$^2$.) These two forces are then resolved into components parallel and orthogonal to the path-tangent at each point, after which the component of the \textit{true force along the tangent} and the component of the \textit{spring force orthogonal to the tangent} are set to zero. Finally, the remaining forces are used to dynamically propagate the system according to some dynamical rule, such as the BFGS or velocity Verlet algorithms. This process is performed iteratively until a self-consistent path is obtained, i.e., the force is converged to within some tolerance (satisfying condition 3). Since NEB identifies a discrete path, condition 2 is not immediately satisfied. However, CI-NEB addresses this by maximizing the energy  of the highest-energy state along the converged NEB path. This is done by setting the path-orthogonal force components to zero and only moving the image along the path.

The optimization of an NEB path based on force convergence criteria can be understood as a constrained geometry optimization of each path image orthogonal to the path. Therefore, to identify the MEP given a discrete initial path between two equilibrium structures, we only require a geometry optimization algorithm capable of operating in the subspace orthogonal to the tangent at each point of the path.

In the case of the surrogate Hessian line-search algorithm, this translates to calculating the surrogate Hessian matrix in the path-orthogonal subspace at each point along an initial guess pathway and performing the line-search within this subspace. Borrowing a DFT-NEB pathway as an initial guess thus permits us to use DMC to obtain a more accurate, corrected MEP. Hereafter, we refer to this as the \textit{subspace optimization} method.
The key steps of subspace optimization for identifying MEPs are the following:
\begin{enumerate}
    \item \textbf{Initialization and parameterization}: We begin with a seed MEP that is obtained using an NEB or CI-NEB calculation within the surrogate theory, i.e., DFT in our case. Suppose this path is given by $\{\mathbf{R}_0\equiv e_A, \mathbf{R}_1, \mathbf{R}_2, \ldots, \mathbf{R}_M, \mathbf{R}_{M+1}\equiv e_B\}$, where $e_A$ and $e_B$ are the initial and final equilibria (e.g., the reactants and products or intermediates for a chemical reaction pathway). Each intermediate image between the initial and final equilibria is then parameterized using the same scheme $\mathrm{F^I}:\mathbb{R}^{3N}\rightarrow \mathbb{R}^d$ that is used to parameterize the equilibrium and transition states to give $\{P_0, P_1, P_2, \ldots, P_M, P_{M+1}\}$, where $P_i\equiv \mathrm{F^I}(\mathbf{R}_i)$.
    \item \textbf{Tangent evaluation}: For each intermediate structure along the pathway, we evaluate the tangent to the path at that structure. There are several ways to accomplish this. Here, we take the vector difference between the current structure and the adjacent structure that is closer to the saddle point. If the current structure is itself the closest structure to the saddle point, we take the mean of the normalized vector differences between the current image and each of its two adjacent images. The key difference with the standard implementation is that we compute the vector differences in structural parameter space rather than in Cartesian coordinates:
    \begin{equation}
        \tau_i = \begin{cases} 
      P_{i+1} - P_i & \text{if } i < i_\textrm{saddle} \\
      \frac{P_{i+1} - P_i}{|P_{i+1} - P_i|} + \frac{P_{i} - P_{i-1}}{|P_i - P_{i-1}|} & \text{if } i = i_\textrm{saddle} \\
      P_{i} - P_{i-1} & \text{if } i > i_\textrm{saddle}
   \end{cases},
    \end{equation}
    where $i_\textrm{saddle}$ is the index of the structure closest to the saddle point.\footnote[1]{We make the simplifying assumption here that the surrogate CI-NEB MEP (correctly) identifies a path whose energy increases monotonically until the saddle point and then decreases monotonically. This allows us to directly index the structure closest to the saddle point using the surrogate MEP energies. If we don't make this assumption, the tangent equation would have to be modified slightly to account for the energy of each image relative to its neighbors' energies in any given iteration of the algorithm.}
    \item \textbf{Subspace vectors and Hessian evaluation}: Having computed the normalized path-tangents, we then use the Gram-Schmidt process to construct unit vectors that are orthogonal to each tangent. For a given tangent $\tau_i$, these unit vectors $\mathbf{v}_i$ then define the path-orthogonal subspace of the overall structural parameter space. At this point, we define secondary mappings $\mathrm{F^{II}}:\mathbb{R}^{d}\rightarrow\mathbb{R}^{d-1}$ that map a point from structural parameter space to the path-orthogonal subspace at each intermediate structure, which is set to be the origin of the subspace. This mapping, $\mathcal{P}_i\equiv \mathrm{F^{II}}(P_i|\mathbf{v}_i)$, is then used as the basis for computing a finite-difference Hessian matrix in the subspace with the surrogate theory. We also construct an inverse mapping $(\mathrm{F^{II}})^{-1}:\mathbb{R}^{d-1}\rightarrow\mathbb{R}^d$ to keep track of line-search optimized structures in the original space of structural parameters.
    \item \textbf{Tolerance optimization and line-search}: Finally, for each intermediate structure, we optimize the line-search parameters $(L_i^\textrm{max}, \sigma_i^\textrm{max})$ in the subspace and perform the usual line-search, as described in Section \ref{sec:eqm_and_saddle}, to locate the local energy minimum within the subspace. We then map each optimized structure, $\mathcal{P}_i^\textrm{opt}$, back to the original space of structural parameters as $(\mathrm{F^{II}})^{-1}(\mathcal{P}_i^\textrm{opt})$, update each intermediate structure to be the new line-search optimum, and repeat from Step 2 onwards.
\end{enumerate}
We summarize the subspace optimization method in Algorithm \ref{alg:subspace_optimization} and show a schematic diagram in Fig. \ref{fig:algorithm_right}. We note that, since the tangents are pre-computed for all the intermediate structures in a given iteration of subspace optimization, the aforementioned operations - corresponding to the inner \textbf{for} loop in Algorithm \ref{alg:subspace_optimization} - can be performed on each structure in parallel.

\begin{algorithm}
\caption{Surrogate Hessian subspace optimization algorithm}\label{alg:subspace_optimization}
\begin{algorithmic}
\State Consider an $N$-atom system with two equilibrium states, $e_A$ and $e_B$, for which an $M$-image MEP is to be determined with structural accuracy $\delta P = \{\delta p_1, \delta p_2, \ldots, \delta p_M\}$.
\State \textbf{start}
\State
$\{\mathbf{R}_0=e_A, \mathbf{R}_1, \mathbf{R}_2, \ldots, \mathbf{R}_M, \mathbf{R}_{M+1}=e_B\} \gets$ \Call{CI-NEB}{$e_A,e_B|M$}
\State $P \gets$ $\{$\Call{$\mathrm{F^I}$}{$\mathbf{R}_0$}, \Call{$\mathrm{F^I}$}{$\mathbf{R}_2$}, $\ldots,$\Call{$\mathrm{F^I}$}{$\mathbf{R}_{M+1}$}$\}$ 
\State \Comment{\Call{$\mathrm{F^I}:\mathbb{R}^{3N}\rightarrow\mathbb{R}^{d}$}{} maps a structure from real space to structural parameter space}
\For{$j\gets 1,2,\ldots,N_\textrm{iter}$} 
\For{$i \gets 1,2,\ldots,M$}
\State $\tau_i \gets $ \Call{Tangent}{$P_i|P_{i-1},P_{i+1}$}
\State \Comment{Computes the path-tangent at image $i$ given its neighboring images}
\State $\mathbf{v}_i \gets$ \Call{GramSchmidt}{$\tau_i$}
\State \Comment{Uses the Gram-Schmidt process to identify $(d-1)$ vectors orthogonal to $\tau_i$}
\State $\mathcal{P}_i \gets$ \Call{$\mathrm{F^{II}}$}{$P_i|\mathbf{v}_i$} 
\State \Comment{\Call{$\mathrm{F^{II}}:\mathbb{R}^{d}\rightarrow\mathbb{R}^{d-1}$}{} maps a structure from structural parameter space to a $(d-1)$-dimensional path-orthogonal subspace}
\State
\State \Comment{Now, we perform the usual surrogate Hessian line-search as described in Ref.~\citenum{original_paper}}
\State $H_i \gets$ \Call{Hessian}{$\mathcal{P}_i$} = $\nabla^2 E_{\textrm{surrogate}}(\mathcal{P}_i)$
\State $(L_i^\textrm{max}, \sigma_i^\textrm{max}) \gets$ \Call{Optimize}{$H_i|\delta P$}
\State $\mathcal{P}_i^\textrm{opt}(\Delta \mathcal{P}_i^\textrm{opt}) \gets$ \Call{LineSearch}{$L_i^\textrm{max},\sigma_i^\textrm{max}$}
\EndFor
\State $\{P_1, P_2, \ldots, P_M\} \gets$ $\{$\Call{$(\mathrm{F^{II}})^{-1}$}{$\mathcal{P}_1^\textrm{opt}$}, \Call{$(\mathrm{F^{II}})^{-1}$}{$\mathcal{P}_2^\textrm{opt}$}, $\ldots$, \Call{$(\mathrm{F^{II}})^{-1}$}{$\mathcal{P}_M^\textrm{opt}$}$\}$
\EndFor
\State \textbf{end}
\end{algorithmic}
\end{algorithm}

Note that while Algorithm \ref{alg:subspace_optimization} depicts running the outer loop for a predefined number of iterations ($N_\textrm{iter}$), this would ideally be replaced with a path convergence-based loop. In practice, however, we find that beginning with a CI-NEB initial guess allows us to obtain accurate paths after just a single iteration, due to which we retain the $\textrm{\textbf{for}}$ loop format shown in the algorithm.

The pathway obtained from this algorithm also presents an alternative approach for locating the saddle point. Instead of computing the surrogate Hessian and initializing the line-search at the highest-energy structure of the DFT-NEB pathway as described in Section \ref{sec:eqm_and_saddle}, we only compute the Hessian at the highest-energy DFT-NEB structure and \textit{initialize the line-search at the highest-energy structure on the DMC subspace optimization pathway}. The utility of this approach is that it permits us to initialize the saddle point line-search at a point on a more accurate (DMC) pathway while only borrowing the search directions (Hessian eigenvectors) from the surrogate theory, thereby providing a more self-consistent protocol for transition state identification. We hereafter refer to this as the self-consistent line-search approach for transition states and discuss the application of this approach in Section \ref{sec:results_eqm_and_saddle}.

\subsubsection{Hybrid DFT-QMC Approach for Thermodynamics and Kinetics}
\label{sec:hybrid_dft_qmc}
Having obtained equilibrium and transition states, and path images, it is desirable to calculate thermodynamic and kinetic quantities using our DMC-optimized geometries and energies. In particular, let us suppose we are interested in computing the free energy changes and equilibrium and rate constants associated with a molecular transformation along an MEP using data from a 0 K ground state \textit{ab initio} theory. This can be done in thermochemistry packages using molecular statistical mechanical principles by evaluating the translational, rotational, and vibrational energies of the structures of interest that compose the enthalpic and entropic contributions to the free energy, given the temperature and the pressure. The equilibrium constant can then be evaluated as

\begin{equation}
    \label{eq:Keq}
    K_\textrm{eq} = \exp \left( -\frac{\Delta G^\circ}{k_B T} \right),
\end{equation}
where $\Delta G^\circ$ is the free energy difference between the final and initial equilibrium states.~\cite{microkinetic_modeling}

Given the transition state, the rate constant for the forward reaction can be obtained using transition state theory as
\begin{equation}
    \label{eq:kf}
    k_\rightarrow = \frac{k_B T}{h}\exp\left(-\frac{\Delta G^{\oplus}}{k_B T}\right),
\end{equation}
where we use $\Delta G^{\oplus}$ to refer to $(\Delta H^{\oplus} - T\Delta S^{\oplus})$.~\cite{microkinetic_modeling} $\Delta S^\oplus$ is the entropy change in transforming from the initial equilibrium state to the transition state, and $\Delta H^{\oplus}$ is the activation energy of the forward reaction. $K_\textrm{eq}$ and $k_\rightarrow$ then automatically determine the backward reaction rate constant as $k_\leftarrow = k_\rightarrow / K_\textrm{eq}$.

Performing these calculations using DMC, however, is not a straightforward task. This is because, in practice, obtaining the vibrational energies requires the evaluation of a finite-difference Hessian matrix in the full $\mathbb{R}^{3N}$ Cartesian representation of the $N$-atom system. In addition to requiring a number of expensive DMC calculations, the intrinsic statistical nature of QMC methods introduces difficulty in  stably and reliably diagonalizing the stochastically sampled Hessian matrix, which is indeed the original motivation for using the surrogate Hessian line-search method.

Surrogate Hessian line-search and subspace optimization are practicable methods due to the - in most cases - correct assumption that the local curvature of the surrogate and stochastic PESs are similar near the structures of interest to a good approximation. It is then reasonable to suggest that we may rely on DFT to provide only the vibrational energies while obtaining the non-vibrational contributions to the enthalpy and entropy from DMC.

In the ideal gas limit, the contributions to the enthalpy may be decomposed as
\begin{equation}
\begin{split}
    H(T) &= E_\textrm{pot} + E_\textrm{ZPE} + \int_0^T C_PdT \\
    &= E_\textrm{pot} + E_\textrm{ZPE} + \int_0^T(k_B + C_{V}^{\textrm{trans}} + C_{V}^{\textrm{rot}} + C_{V}^{\textrm{vib}} + C_{V}^{\textrm{elec}})dT,
    \end{split}
    \label{eq:full_enthalpy_original}
\end{equation}
where $E_\textrm{pot}$ is the potential energy, $E_\textrm{ZPE}$ is the zero-point energy, and $C_P$ is the heat capacity at constant pressure, which can be further broken down into the sum of the translational ($C_{V}^{\textrm{trans}}$), rotational ($C_{V}^{\textrm{rot}}$), vibrational ($C_{V}^{\textrm{vib}}$), and electronic ($C_{V}^{\textrm{elec}}$) heat capacities at constant volume.~\cite{cramer_computational_chemistry} We suppose that the zero-point energy and the vibrational component of the heat capacity can be obtained from DFT. Ignoring the electronic heat capacity, it is now clear that the potential energy, $E_\textrm{pot}$, is the only other component of the enthalpy that is sensitive to the choice of \textit{ab initio} method used to optimize the structure and calculate its energy. We may then make an approximation to the enthalpy that leverages the improved accuracy of DMC over DFT:
\begin{equation}
\begin{split}
    H_\textrm{hybrid}(T) = \underbrace{E_\textrm{pot}^\textrm{DMC}}_{\text{calculated at } \langle\textbf{R}_\textrm{DMC}^*\rangle
    } + \underbrace{E_\textrm{ZPE}^\textrm{DFT} + \int_0^T C_{V,\textrm{vib}}dT}_{\text{calculated at } \textbf{R}_\textrm{DFT}^*} \\
    +
    \int_0^T(k_B + C_{V}^{\textrm{trans}} + C_{V}^{\textrm{rot}})dT,
    \label{eq:H_hybrid_enthalpy}
\end{split}
\end{equation}
with the statistical error in the DMC potential energy now affecting the enthalpy directly as
\begin{equation}
    \sigma(H_\textrm{hybrid}(T)) = \sigma(E_\textrm{pot}^\textrm{DMC}),
\end{equation}
which is just the standard error of the DMC energy.

It is important to note that $E_\textrm{pot}^\textrm{DMC}$ and $(E_\textrm{ZPE}^\textrm{DFT}, C_V^\textrm{vib}|_\textrm{DFT})$ are calculated using two slightly different geometries in Eq. \ref{eq:H_hybrid_enthalpy}. $E_\textrm{pot}^\textrm{DMC}$ is taken to be the potential energy of the mean surrogate Hessian-optimized DMC geometry ($\langle\textbf{R}_\textrm{DMC}^*\rangle$), while the vibrational energies and all ensuing contributions are calculated using the DFT-optimized geometry ($\textbf{R}_\textrm{DFT}^*$). This is crucial for the following reason. Assume we are interested in an equilibrium structure. Then, the local features of the PES near this structure will only be reflected in the DFT Hessian matrix if the matrix is calculated around the DFT-equilibrium structure. Since this may be slightly displaced from the DMC-equilibrium structure, constructing a DFT Hessian matrix centered at the latter may capture the curvature of an irrelevant region of the \textit{DFT} PES. Instead, we combine the curvature (Hessian matrix) near the equilibrium of the surrogate, i.e., DFT, PES with the potential energy near the equilibrium of the more accurate, stochastic PES (DMC). The same principle applies to saddle point structures. This strategy is valid because the notion of a critical point within some sufficiently local region of a PES is independent of the exact structural parameters to which it corresponds and the specific level-of-theory that was used to obtain it. This allows us to combine two slightly different structures optimized using different methods because they theoretically refer to the same region of the PES.

Now, let us turn our attention to the entropy, which can be decomposed as
\begin{equation}
    S(T, P) = S_\textrm{trans}(T,P^\circ) + S_\textrm{rot}(T) + S_\textrm{vib}(T) + S_\textrm{elec} - k_B\ln\frac{P}{P^\circ},
    \label{eq:full_entropy_original}
\end{equation}
where $P^\circ$ is the pressure at the standard state and $S_\textrm{trans}$, $S_\textrm{rot}$, $S_\textrm{vib}$, and $S_\textrm{elec}$ refer to the translational, rotational, vibrational, and electronic entropies respectively.~\cite{cramer_computational_chemistry} Again, $S_\textrm{trans}$ and $S_\textrm{elec}$ are unaffected by slight differences between structures optimized by different \textit{ab initio} methods. As earlier, we can rely on DFT to compute $S_\textrm{vib}$. Computing $S_\textrm{rot}$ analytically, however, is a non-trivial task since it involves the calculation of molecular moments-of-inertia - given the uncertainty in the structural parameters of a DMC-optimized molecule, it is challenging to predict the propagation of these errors to the moments-of-inertia and the rotational entropy. Not only are the errors in structural parameters correlated, but the propagation of this error also becomes hard to generalize for differently parameterized systems.

To overcome this, we employ a Monte Carlo scheme to estimate the rotational entropy stochastically. First, we sample a large number of structures using the mean and variance of the DMC-optimized structural parameters. After converting these structures back to their Cartesian representation, we calculate the rotational entropy for each of these sampled structures independently, and finally report the average entropy and variance:
\begin{equation}
\begin{split}
    S_{\textrm{hybrid}}(T, P) = \underbrace{S_\textrm{rot}(T)}_{\text{Monte Carlo sampling around } \langle\textbf{R}_\textrm{DMC}^*\rangle} + \underbrace{S_\textrm{vib}(T)}_{\text{calculated at } \textbf{R}_\textrm{DFT}^*} \\
    + S_\textrm{trans}(T,P^\circ) + S_\textrm{elec} - k_B\ln\frac{P}{P^\circ},
    \label{eq:S_hybrid_entropy}
    \end{split}
\end{equation}
so that
\begin{equation}
    \sigma(S_\textrm{hybrid}(T)) = \sqrt{\frac{\sum_{i=1}^{N_\textrm{MC}}\left(S_\textrm{rot}^i(T)-\langle S_\textrm{rot}(T)\rangle\right)^2}{N_\textrm{MC}}},
\end{equation}
where $S_\textrm{rot}^i(T)$ is the rotational entropy of a structural sample $i$ around $\langle\mathbf{R}_\textrm{DMC}^*\rangle$ and $N_\textrm{MC}$ is the number of Monte Carlo samples.

From here, it is easy to calculate the free energy changes and equilibrium and rate constants from Eqs. \ref{eq:Keq} and \ref{eq:kf}.

\subsection{Illustrative Test Cases}
\label{sec:testcases}
\begin{figure}[!ht]
    \centering
    \begin{subfigure}[b]{0.7\textwidth}{\includegraphics[width=\textwidth]{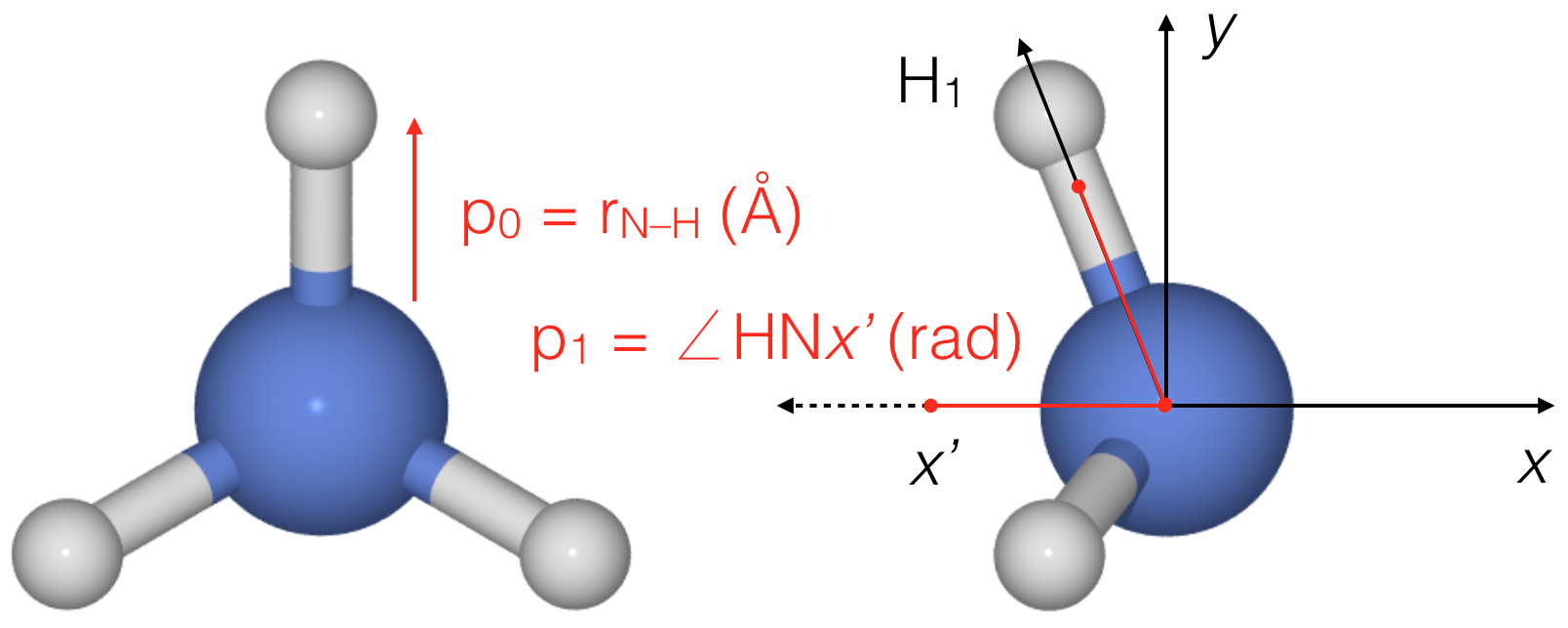}}
    \caption{}
    \label{fig:NH3_parameterization}
    \end{subfigure}
    \vspace{5mm}
    
    \hrulefill
    
    \vspace{8mm}
    \begin{subfigure}[b]{0.7\textwidth}{\includegraphics[width=\textwidth]{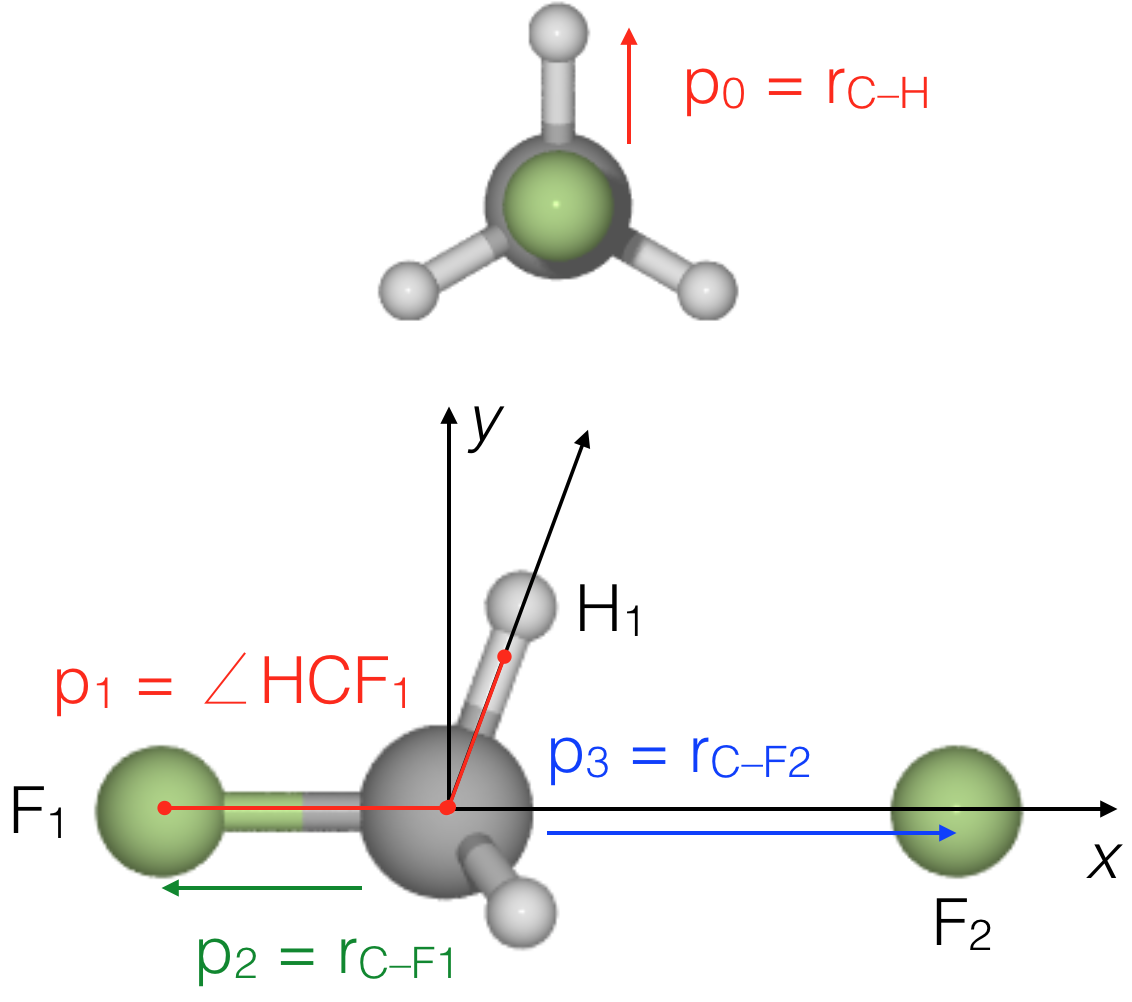}}
    \caption{}
    \label{fig:SN2_parameterization}
    \end{subfigure}
    \caption{Structural parameterization of (a) the NH$_3$ inversion system, and (b) the S$_\textrm{N}$2 reaction system. All bond lengths are measured in Å and bond angles in radians.}
    \label{fig:parameterization}
\end{figure}

In this work, we focus on two illustrative chemical processes as test cases: (1) ammonia (NH$_3$) inversion and (2) the nucleophilic substitution (S$_\textrm{N}$2) reaction F$^-$ + CH$_3$F $\rightarrow$ FCH$_3$ + F$^-$. Ammonia inversion is a process that occurs $\sim$30 billion times per second\cite{more_is_different} at room temperature in which the three hydrogen atoms that form a trigonal pyramid around nitrogen flip from one side of the N to the other. In the S$_\textrm{N}$2 reaction, one fluorine substitutes for another in CH$_3$F in a concerted (hence the 2) fashion, resulting in an inversion of the methyl group to accommodate the bond rearrangement. This and other S$_\textrm{N}$2-type systems have been studied extensively using \textit{ab initio} methods; see, for example, Ref.~\citenum{reactions}. We now describe examples of minimal structural parameterizations of the NH$_3$ inversion and S$_\textrm{N}$2 reaction systems that describe the transformations (inversion and bond rearrangement, respectively) that these systems undergo.
\\
We define the geometry of the NH$_3$ system using two structural parameters (Fig. \ref{fig:NH3_parameterization}): the mean N-H bond length in Å ($p_0$) and the mean angle in radians ($p_1$) formed by the N-H bonds with the negative $x$-axis (labeled $x'$). Without loss of generality, we assume that the centroid of the three H atoms lies on the $x$-axis in all configurations during the inversion of the molecule.
We set the N atom to always coincide with the origin and constrain one of the H atoms (labeled H$_1$) to lie in the $x-y$ plane. We similarly parameterize the S$_\textrm{N}$2 reaction system (Fig. \ref{fig:SN2_parameterization}) using the mean C-H bond length ($p_0$) and the mean $\measuredangle$HC$x'$ bond angle ($p_1$). Additionally, we include the two C-F bond lengths, C-F$_1$ ($p_2$) and C-F$_2$ ($p_3$). We set the C atom to coincide with the origin, the H$_1$ atom to lie in the $x-y$ plane, and the C-F$_1$ bond to lie along the $x'$ axis. It follows that $\measuredangle$HC$x'$ = $\measuredangle$HCF$_1$. Explicit expressions for both parameterizations are provided in Section \ref{supporting:mapping} of the Supporting Information.

In both systems, the initial and final equilibrium geometries are identical and share $C_{3v}$ symmetry. However, grounding the parameterization in a Cartesian reference frame enables us to distinguish between the initial and final states in our analysis.

\subsection{Computational Details}
\label{sec:computational_details}
For the NH$_3$ system, we set error tolerances of 0.01 Å for the bond length ($\delta p_0$) and 0.05 rad for the bond angle ($\delta p_1$). For the S$_\textrm{N}$2 reaction system, we set the following tolerances: $(\delta p_0 = 0.01$ Å$, \delta p_1 = 0.05$ rad$, \delta p_2 = 0.05$ Å, and $\delta p_3 = 0.05$ Å$)$. We choose higher tolerances for the two C-F bond lengths because they vary to a greater extent over the course of the reaction, which enables their variation to be captured accurately with lower precision and hence lower computational cost.

We perform all DFT and Coupled Cluster with Single and Double excitations (CCSD) calculations in PySCF.~\cite{pyscf_1, pyscf_2} We use correlation-consistent effective core potentials (ccECPs) and basis sets (cc-pVTZ-ccECP).~\cite{ccecp, ccecp_2} We use the PBE\cite{pbe_functional} functional for our DFT calculations and use identical DFT settings for computing energies, trial wave functions, and DFT-NEB paths.

For each QMC calculation, following DFT trial wave function generation, we optimize one- and two-body Jastrow factors via 8 optimization cycles of 100,000 samples. We then further optimize the trial wave function with a VMC calculation by taking 1000 samples. Finally, we perform a DMC calculation with a timestep of 0.005 a.u. and run it for the estimated number of steps required to compute the energies within the stipulated tolerances. We note that translating error tolerances in structural parameters ($\delta p$) into the number of DMC steps required to saturate these tolerances involves a series of approximations (see Ref.~\citenum{original_paper}) and can result in higher or lower structural errors than intended. To mitigate this, we extend the number of DMC steps in some of our calculations by a factor of 16 or 64 (structure-wise values can be found in Section \ref{supporting:extension}). We perform all QMC calculations using QMCPACK\cite{qmcpack_1, qmcpack_2} and assemble the DFT-VMC-DMC pipeline using the Nexus workflow management system.~\cite{nexus}

We perform all NEB calculations in the Atomic Simulation Environment (ASE)\cite{ase} using the BFGS\cite{ase} algorithm and a force tolerance of 0.01 eV/Å. For the NH$_3$ system, we include 3 intermediate images between the initial and final equilibrium states, whereas for the S$_\textrm{N}$2 system, we include 5 intermediate images. To locate the saddle point, we perform a climbing image calculation on the structure with the highest energy along the MEP. We also use the BFGS algorithm for DFT and CCSD equilibrium geometry optimization. In this case, we set the force tolerance to 0.001 eV/Å.

\section{Results and Discussion}
\label{sec:results_and_discussion}
\subsection{Equilibrium and Transition State Geometries}
\label{sec:results_eqm_and_saddle}
First, we report the results of the surrogate Hessian line-search method applied to the optimization of the equilibrium and transition state geometries of our two test systems. Tables \ref{tab:NH3_eqm_and_saddle} (NH$_3$) and \ref{tab:SN2_eqm_and_saddle} (S$_\textrm{N}$2) compare the DFT and CCSD structural relaxations and CI-NEB results for the transition states to those obtained using the surrogate Hessian line-search approach using DMC as described in Section \ref{sec:eqm_and_saddle} for the NH$_3$ inversion and S$_\textrm{N}$2 reaction systems. Additionally, to illustrate that our approaches are self-consistent, we include the results for saddle point identification using the self-consistent approach described in Section \ref{sec:subspace_optimization}. For this approach, we compute the line-search directions as the eigenvectors of the surrogate Hessian at the DFT-NEB saddle point, while the line-search itself is initialized at a high-energy image along the DMC MEP obtained using subspace optimization. Given the structural symmetries of the NH$_3$ and S$_\textrm{N}$2 systems and the odd number of intermediate images we have considered here, the highest-energy image already coincides with the saddle point in both cases. Thus, to make for more challenging test cases, we instead seed the self-consistent saddle point line-search at a point adjacent to the highest-energy image along each path. This corresponds to DMC path image 3 in Fig. \ref{fig:NH3_E_vs_rc_GPR} for the NH$_3$ case, and DMC path image 2 in Fig. \ref{fig:SN2_E_vs_rc_GPR} for the S$_\textrm{N}2$ case.

\begin{table}
\centering
\begin{tabular}{|c|c|c|c|c|c|c|}
\hline
    \multirow{2}{*}{} & \multicolumn{2}{|c|}{\textbf{Equilibrium A}} & \multicolumn{2}{|c|}{\textbf{Transition State}} & \multicolumn{2}{|c|}{\textbf{Equilibrium B}} \\ \cline{2-7}
     & $p_0$ (Å) & $p_1$ (rad) & $p_0$ (Å) & $p_1$ (rad) & $p_0$ (Å) & $p_1$ (rad) \\ \hline
\textbf{DFT} & 1.019 & 1.958 & 0.998 & 1.571 & 1.019 & 1.184 \\ \hline
\textbf{CCSD} & 1.010 & 1.959 & 0.990 & 1.571 & 1.010 & 1.183 \\ \hline
\textbf{DMC (LS)} & 1.006(1) & 1.957(5) & 0.991(2) & 1.569(14) & 1.007(2) & 1.202(10)
\\ \hline
\textbf{DMC (LS-SC)} & - & - & 0.992(2) & 1.574(34) & - & -
\\ \hline
\end{tabular}
\caption{NH$_3$ structural parameters obtained using:
[Row 1] DFT structural relaxation for the equilibrium states and CI-NEB for the transition state; [Row 2] CCSD structural relaxation for the equilibrium states and CI-NEB for the transition state; [Row 3] DMC surrogate Hessian line-search; and [Row 4] DMC self-consistent saddle point line-search initialized at path image 3 in Fig. \ref{fig:NH3_E_vs_rc_GPR}.}
\label{tab:NH3_eqm_and_saddle}
\end{table}

\begin{table}
\centering
\begin{tabular}{|c|c|c|c|c|}
\hline
    \multirow{2}{*}{} & \multicolumn{4}{|c|}{\textbf{Equilibrium A}} \\ \cline{2-5}
    & $p_0$ (Å) & $p_1$ (rad) & $p_2$ (Å) & $p_3$ (Å) \\ \hline
    \textbf{DFT} & 1.094 & 1.218 & 2.575 & 1.434 \\ \hline
    \textbf{CCSD} & 1.079 & 1.243 & 2.575 & 1.434 \\ \hline
    \textbf{DMC (LS)} & 1.079(3) & 1.244(16) & 2.580(17) & 1.412(11) \\ \hline
\hline
    \multirow{2}{*}{} & \multicolumn{4}{|c|}{\textbf{Transition State}} \\ \cline{2-5}
    & $p_0$ (Å) & $p_1$ (rad) & $p_2$ (Å) & $p_3$ (Å) \\ \hline
    \textbf{DFT} & 1.076 & 1.571 & 1.863 & 1.863 \\ \hline
    \textbf{CCSD} & 1.063 & 1.571 & 1.849 & 1.849 \\ \hline
    \textbf{DMC (LS)} & 1.066(3) & 1.568(14) & 1.824(13) & 1.816(14) \\ \hline
    \textbf{DMC (LS-SC)} & 1.068(6) & 1.565(20) & 1.851(30) & 1.794(28) \\ \hline
\hline
    \multirow{2}{*}{} & \multicolumn{4}{|c|}{\textbf{Equilibrium B}} \\ \cline{2-5}
    & $p_0$ (Å) & $p_1$ (rad) & $p_2$ (Å) & $p_3$ (Å) \\ \hline
    \textbf{DFT} & 1.094 & 1.923 & 1.434 & 2.575 \\ \hline
    \textbf{CCSD} & 1.079 & 1.899 & 1.434 & 2.575 \\ \hline
    \textbf{DMC (LS)} & 1.078(2) & 1.894(8) & 1.412(10) & 2.580(14) \\ \hline
\end{tabular}
\caption{S$_\textrm{N}$2 reaction structural parameters obtained using:
[Row 1] DFT structural relaxation for equilibrium states and CI-NEB for the transition state; [Row 2] CCSD structural relaxation for equilibrium states and CI-NEB for the transition state; [Row 3] DMC surrogate Hessian line-search; [Row 4 in Transition State block] DMC self-consistent saddle point line-search initialized at path image 2 in Fig. \ref{fig:SN2_E_vs_rc_GPR}.}
\label{tab:SN2_eqm_and_saddle}
\end{table}

From Tables \ref{tab:NH3_eqm_and_saddle} and \ref{tab:SN2_eqm_and_saddle}, it is clear that the DMC line-search yields geometries that are in excellent agreement with CCSD geometries obtained using BFGS structural relaxation, with differences in bond lengths of less than $\sim$0.02 Å and differences in bond angles of less than $\sim$0.02 rad (or 1.15$^\circ$). Additionally, agreement between the direct DMC 
line-search (DMC (LS)) and self-consistent DMC line-search (DMC (LS-SC)), within statistical variation, indicates the capability of the surrogate Hessian approach to determine transition state structures both in the presence and absence of initial guess transition state geometries obtained from surrogate theories such as DFT. We note that, while there are slightly larger ($\sim$0.05 Å) differences between DMC and CCSD in the two C-F bond lengths ($p_2$ and $p_3$) in the S$_\textrm{N}$2 example, particularly in the transition state structure, these are a consequence of the very low sensitivity of the potential energy to these structural parameters. Since the surrogate Hessian line-search method is based on non-linearly combining structural tolerances into an energetic tolerance, locating a critical point to within some energetic tolerance can incur - based on how strongly (or weakly) they are coupled - commensurate variability in the structural parameters. This will become clearer when we compare the energy barriers along the MEP obtained by each method. For a more detailed analysis of the coupling of energetic and structural tolerances, we refer the reader to Ref. \citenum{original_paper}.

\subsection{Minimum-Energy Pathways}
\label{sec:results_mep}
We now discuss the use of the subspace optimization algorithm to determine the MEPs of the NH$_3$ inversion and the S$_\textrm{N}$2 reaction systems. We begin by defining two equilibrium states, i.e., the DFT-optimized states labeled Equilibrium A and Equilibrium B in Tables \ref{tab:NH3_eqm_and_saddle} and \ref{tab:SN2_eqm_and_saddle}. For example, in the NH$_3$ system, Equilibrium B refers to the molecule shown in Fig. \ref{fig:NH3_parameterization} with the angle $\measuredangle$HN$x'$ being acute, while Equilibrium A refers to the inverted structure of this molecule, where the angle $\measuredangle$HN$x'$ becomes obtuse. In the S$_\textrm{N}$2 case, both the angle $\measuredangle$HCF$_1$ and the C-F$_1$ and C-F$_2$ bond lengths change in going from Equilibrium A to Equilibrium B. For consistency of MEP comparison between the various methods, we use these DFT-optimized configurations as the fixed endpoints for all MEP calculations.

\begin{figure}[ht]
    \centering
    \includegraphics[width=0.7\textwidth]{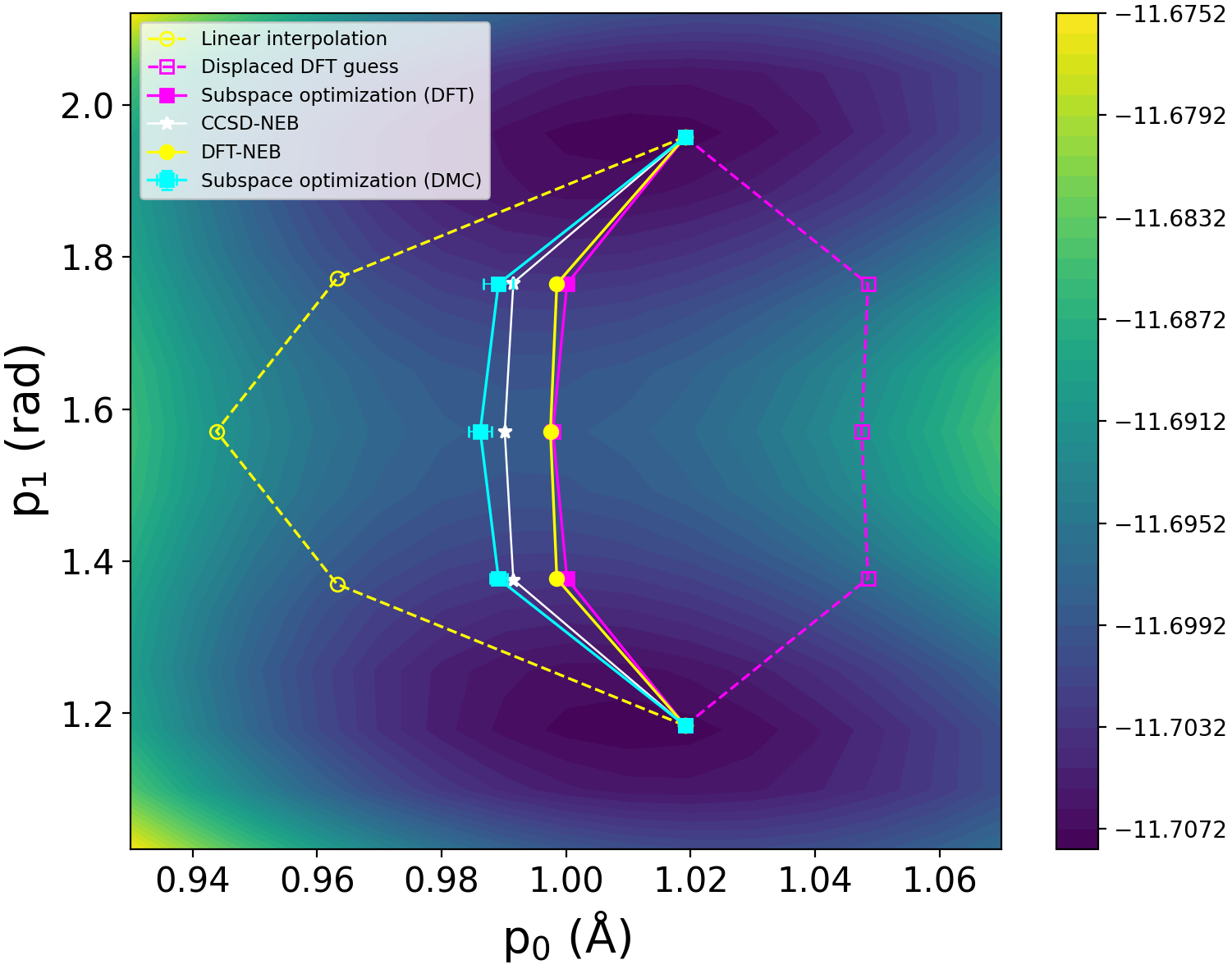}
    \caption{MEPs for NH$_3$ inversion obtained using subspace optimization and NEB for different methods overlaid on a CCSD energy heat map. Energies are measured in eV. The linear interpolation path is used as an initial guess for the DFT-NEB and CCSD-NEB calculations. The displaced DFT guess path (displaced from the fully converged DFT-NEB path) is used to seed the DFT subspace optimization calculation. The \textit{x}- and \textit{y}-axes indicate the NH$_3$ structural parameters, $p_0$ and $p_1$, respectively.}
    \label{fig:NH3_subspace_optimization}
\end{figure}

Starting with a linear interpolation between these two equilibria, we independently perform NEB calculations using DFT and CCSD. Then, we use the DFT-NEB states to compute the subspace surrogate Hessian and its eigenvectors for each intermediate structure along the pathway. To compare a DFT-based surrogate Hessian subspace optimization with the DFT-NEB pathway, we displace the DFT-NEB images by a small fixed amount of 0.05 units along each of the subspace eigenvectors. (It is difficult to associate this shift with a particular unit, such as Å or radian, because the subspace is constructed by taking combinations of vectors along the individual structural parameters, due to which a certain vector in the path-orthogonal subspace has a component along more than one structural parameter with no unique unit of measurement.) We then use these displaced states to initialize the DFT-based subspace optimization. We then proceed to identify the DMC MEP by performing a DMC-based subspace optimization using the DFT-NEB pathway as an initial guess with no displacement along the subspace eigenvectors. To illustrate the differences between these pathways, we plot the NH$_3$ inversion pathway in structural parameter space overlaid on an energy heat map calculated using a fine grid of CCSD calculations in Fig. \ref{fig:NH3_subspace_optimization}.

It is clear from Fig. \ref{fig:NH3_subspace_optimization} that the subspace optimization algorithm correctly identifies the MEP due to the close agreement between the DFT-NEB and DFT-subspace optimization pathways. Furthermore, the DMC subspace optimization pathway shows better structural agreement with the CCSD-NEB pathway than the DFT-based pathways, indicating the potential for more accurate (compared to DFT) and inexpensive (compared to CCSD) MEP identification using surrogate Hessian subspace optimization.

At this point, we note that, despite good agreement with NEB, subspace optimization and NEB are not strictly equivalent algorithms and therefore cannot always be expected to yield fully identical MEPs. The following are a few notable distinctions between typical implementations of NEB and subspace optimization:
\begin{enumerate}
    \item \textbf{Tangent calculation}: During an NEB calculation, tangents to the MEP at each intermediate structure are calculated using vector differences between adjacent structures in Cartesian coordinates. Subspace optimization instead computes these tangents in the space of structural parameters. Since the operations that map Cartesian coordinates to reduced structural parameters may be non-distributive over vector differences between structures, the tangents - and thereby, the path-orthogonal directions - computed using NEB and subspace optimization may not be identical. Neither approach is necessarily ``more correct" since the choice of coordinates is a matter of representational convenience. An alternative representation-invariant approach (not addressed here) would be to construct an internal coordinate representation that accounts for the implicit metric change that occurs when computing tangents at points located in different regions of the PES\cite{geodesic_interpolation}.
    \item \textbf{Hessian evaluation}: In NEB calculations that use force-based dynamical algorithms such as BFGS to propagate the chain of states toward the MEP, the Hessian matrix is evaluated and updated at each step of the dynamics. This results in the possibility of multiple Hessian evaluations even within a single NEB iteration. By contrast, the subspace optimization algorithm is currently implemented to only perform a single evaluation of the subspace Hessian in a given iteration, which is then used to perform multiple line-search iterations. It is for this reason that providing good starting structures for Hessian evaluation is necessary for this algorithm to operate efficiently. While multiple Hessian evaluations can be implemented, we do not include this in our approach.
    \item \textbf{Use of springs}: Since NEB is based on the resolution of forces along and orthogonal to path-tangents, it is reasonable to introduce artificial spring forces to maintain an even spacing of images along the path. However, since such resolution of components is not possible in a purely energy-based optimization scheme, subspace optimization does not include spring forces and instead relies on evenly spaced starting images.
    \item \textbf{Termination/convergence criterion}: Implementations of NEB are typically based on force-based convergence, where the method is set to iterate until the force on each atom falls below some threshold. Subspace optimization, on the other hand, may be implemented using a path-based convergence criterion, where the structural difference between consecutive paths must fall below a certain threshold for convergence.
\end{enumerate}

\begin{figure}[ht]
    \centering
    {\includegraphics[width=0.8\textwidth]{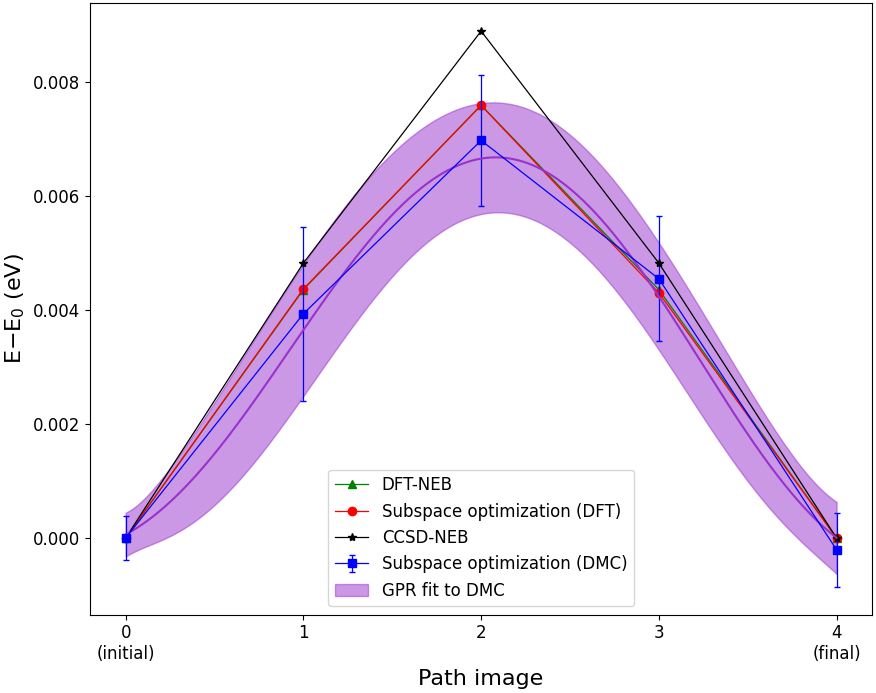}}
    \caption{Energy vs. NH$_3$ inversion path image for different methods. The DFT-NEB and DFT subspace optimization points overlap almost completely.}
    \label{fig:NH3_E_vs_rc_GPR}
\end{figure}

Having obtained MEPs from each of the methods, we now consider the energy barriers along the MEPs obtained with each method. We plot this for the NH$_3$ system in Fig. \ref{fig:NH3_E_vs_rc_GPR}, with the inclusion of a Gaussian process regression fit to the DMC data points to depict the smooth variation of the mean energy and standard energy error along the MEP. It must be noted that, while we plot each energy barrier with a common $x$-axis labeled `Path image,' these do not correspond to the same reaction coordinate, since the MEP obtained using each method is slightly different from the rest.

It is clear from Fig. \ref{fig:NH3_E_vs_rc_GPR} that DMC-based subspace optimization is capable of resolving energy barriers to within a few meV and is in very good agreement with the CCSD-NEB energy barrier. We also note that in Fig. \ref{fig:NH3_E_vs_rc_GPR}, the DFT-NEB and DFT subspace optimization barriers are so close that the DFT-NEB data points are not visible.

\begin{figure}[ht]
    \centering
    \begin{subfigure}[b]{0.49\textwidth}{\includegraphics[width=\textwidth]{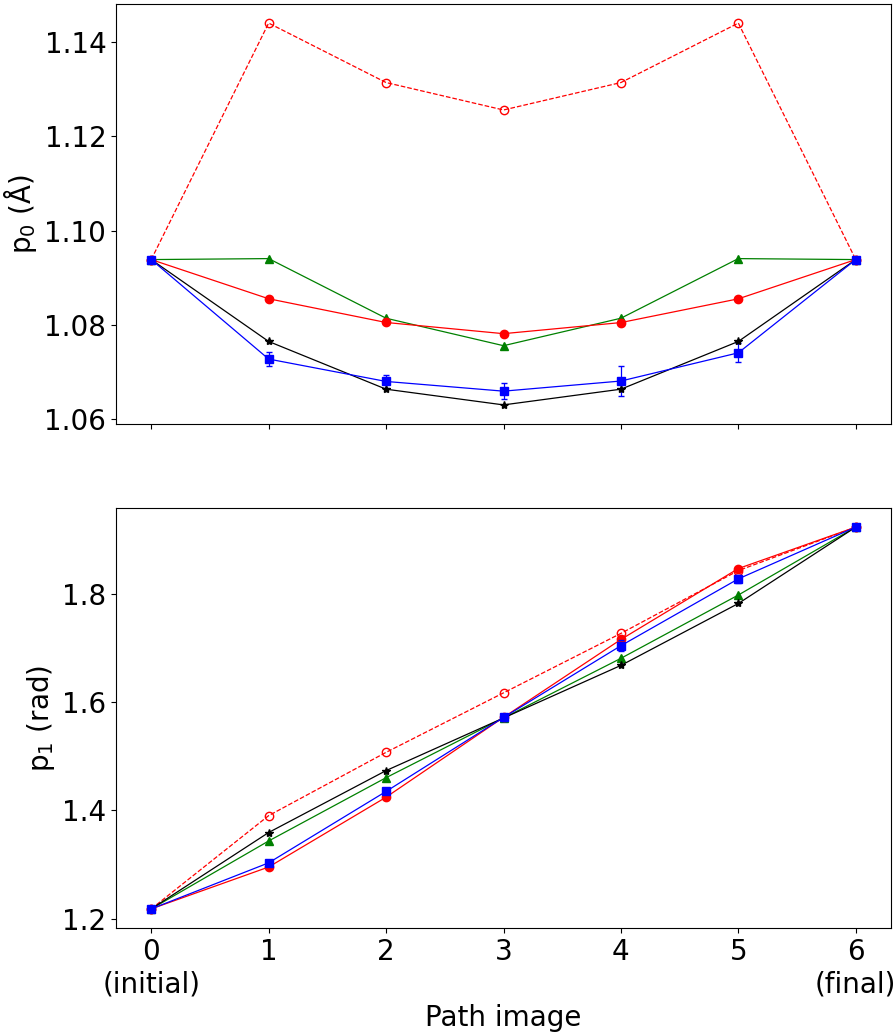}}
    \label{fig:p01_vs_rc_GPR}
    \end{subfigure}
    \hfill
    \begin{subfigure}[b]{0.49\textwidth}{\includegraphics[width=\textwidth]{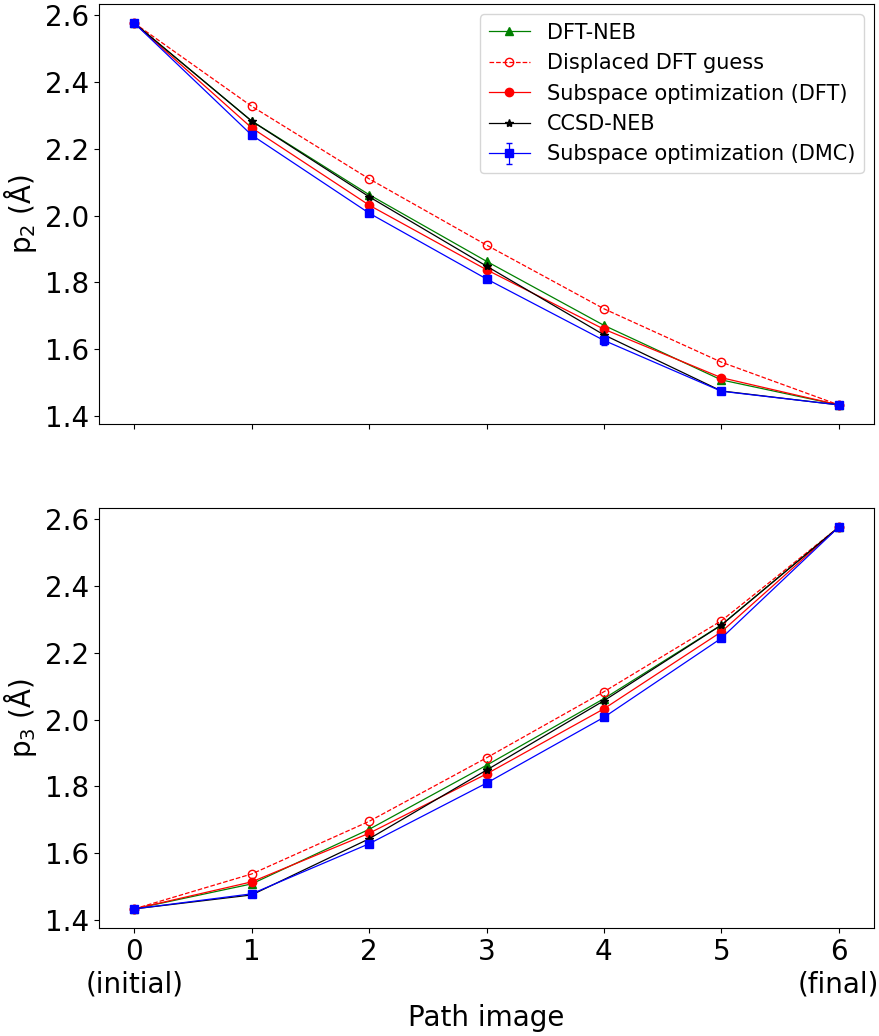}}
    \label{fig:p23_vs_rc_GPR}
    \end{subfigure}
    \caption{Structural parameters vs. path image of the S$_\textrm{N}$2 reaction system for different methods (see top right panel for legend). The linear interpolation path is used as an initial guess for the DFT-NEB and CCSD-NEB calculations. The displaced DFT guess path (displaced from the fully converged DFT-NEB path) is used to seed the DFT subspace optimization calculation.}
    \label{fig:p_vs_rc}
\end{figure}

\begin{figure}[ht]
    \centering
    {\includegraphics[width=0.8
\textwidth]{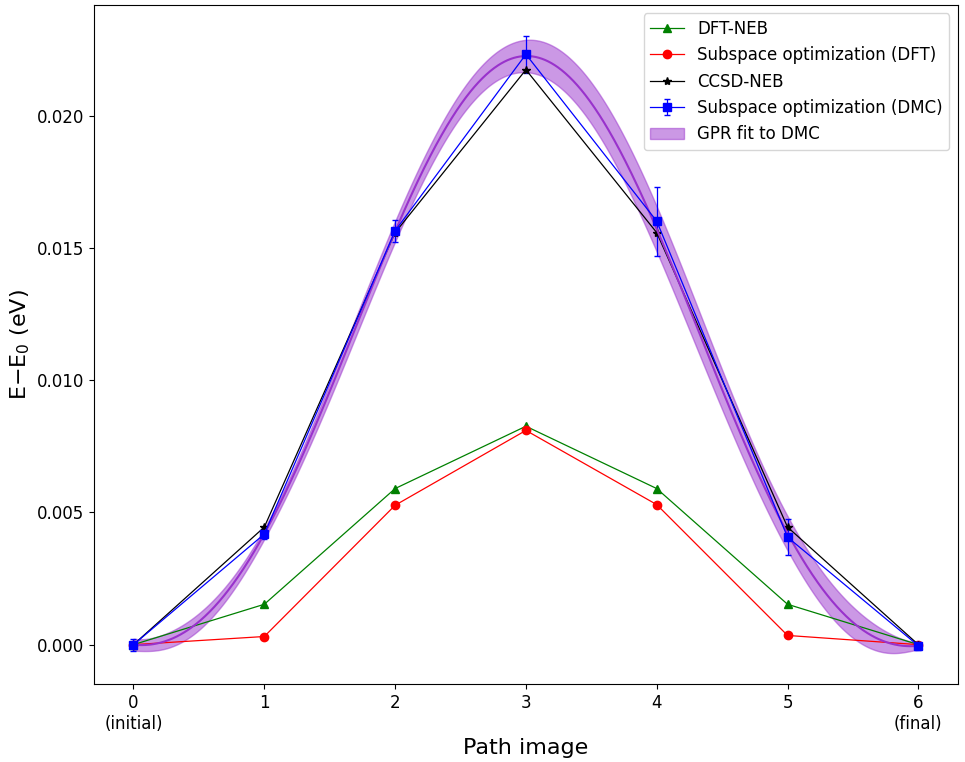}}
    \caption{Energy vs. S$_
    \textrm{N}$2 reaction path image for different methods.}
    \label{fig:SN2_E_vs_rc_GPR}
\end{figure}

We now turn to the S$_\textrm{N}$2 reaction system, which possesses more degrees of freedom and is expected to exhibit greater electron correlation, leading to more significant differences between the MEPs. Since it is difficult to show the variation of all structural parameters in the same plot, we show the variation of each structural parameter with respect to the path image in Fig. \ref{fig:p_vs_rc}.
Again, it is evident that subspace optimization accurately captures the trends in the variation of structural parameters across the MEP using both DFT and DMC. We compare the energy barriers obtained by the different methods in Fig. \ref{fig:SN2_E_vs_rc_GPR}. Here, we see that DMC subspace optimization is in excellent agreement with CCSD along the MEP, while the DFT-NEB and DFT subspace optimization differ by less than 2.5 meV, or $\sim$ 0.1 mHa, from each other. Note that the DFT and DMC MEP barrier heights differ by nearly 20 meV from one another, illustrating that correlation assumes a more significant role in this test case. 

We note that, due to the distinctions between NEB and subspace optimization mentioned above, the two methods do not yield perfectly identical MEP structures at the DFT level. In our method, the structures along the MEP are determined as a consequence of path-orthogonal energy optimization. Given that several structures can be energetically very similar, the energy optimization algorithm may rightly identify any one of these structures as lying on the MEP, although it may not necessarily coincide with the structure identified by a force-minimization algorithm such as NEB.
Due to these practically insignificant energy differences, paths obtained by both methods can be considered to consist of representative points along the MEP despite small structural dissimilarities.

\subsection{Free Energies and Kinetic Parameters}
\label{sec:results_hybrid_dft_qmc}
We now discuss the evaluation of thermodynamic and kinetic quantities obtained using the hybrid DFT-QMC approach described in Section \ref{sec:hybrid_dft_qmc}. In particular, we report the free energy difference between the saddle point and equilibrium structures ($\Delta G^\oplus$ in Eq. \ref{eq:Keq}) and between the two equilibrium states ($\Delta G^\circ$ in Eq. \ref{eq:kf}). We also report the associated kinetic parameters, $k_\rightarrow$ and $K_\textrm{eq}$. We benchmark these results against DFT and CCSD calculations using the standard approach, in which all energetic and entropic contributions are computed deterministically using the same geometry.~\cite{ase,cramer_computational_chemistry}

During the Monte Carlo structure sampling to estimate the rotational entropy ($S_\textrm{rot}(T)$ in Eq. \ref{eq:S_hybrid_entropy}), we sample 2500 structures each for the saddle point and equilibrium states of NH$_3$ using the mean and standard error of the structural parameters identified using DMC in Table \ref{tab:NH3_eqm_and_saddle}. Similarly, in the case of the S$_\textrm{N}$2 reaction, we sample 10,000 structures around the mean and standard error values shown in Table \ref{tab:SN2_eqm_and_saddle}.

\begin{table}
\centering
\begin{tabular}{|c|c|c|c|c|}
\hline
     & $\Delta G^{\oplus}$ (eV) & $k_\rightarrow(\times 10^{16}$ s$^{-1})$ & $\Delta G^\circ$ (eV) & $K_\textrm{eq}$ \\ \hline
\textbf{DFT} & -0.199 & 1.436 & 0.000 & 1.000 \\ \hline
\textbf{Hybrid DFT-QMC} & -0.200(0) & 1.483(24) & -0.000(1) & 1.018(33) \\ \hline
\textbf{CCSD} & -0.222 & 3.514 & 0.000 & 1.000 \\ \hline
\end{tabular}
\caption{Free energy differences and equilibrium and rate constants for the NH$_3$ inversion system.}
\label{tab:NH3_kinetics}
\end{table}

\begin{table}
\centering
\begin{tabular}{|c|c|c|c|c|}
\hline
     & $\Delta G^{\oplus}$ (eV) & $k_\rightarrow(\times 10^{12}$ s$^{-1})$ & $\Delta G^\circ$ (eV) & $K_\textrm{eq}$ \\ \hline
\textbf{DFT} & 0.004 & 5.317 & 0.000 & 1.000 \\ \hline
\textbf{Hybrid DFT-QMC} & 0.019(1) & 2.918(122) & 0.000(0) & 0.989(19) \\ \hline
\textbf{CCSD} & 0.020 & 2.852 & 0.000 & 1.000 \\ \hline
\end{tabular}
\caption{Free energy differences and equilibrium and rate constants for the S$_\textrm{N}$2 reaction system.}
\label{tab:SN2_kinetics}
\end{table}

The free energy differences and equilibrium and rate constants are shown in Table \ref{tab:NH3_kinetics} for the NH$_3$ inversion system and in Table \ref{tab:SN2_kinetics} for the S$_\textrm{N}$2 reaction system. Tables \ref{tab:NH3_thermodynamic_breakdown} and \ref{tab:SN2_thermodynamic_breakdown} show the free energy contributions of the various \textit{ab initio} method-dependent terms in Eqs. \ref{eq:full_enthalpy_original}, \ref{eq:H_hybrid_enthalpy}, \ref{eq:full_entropy_original}, and \ref{eq:S_hybrid_entropy}. As can be seen in Table \ref{tab:NH3_thermodynamic_breakdown}, the dominant contribution to $\Delta G^\oplus$ in the NH$_3$ system is the \textit{difference in vibrational entropies} of the transition state and the equilibrium state, which DFT underestimates by $\sim$0.022 eV compared to CCSD. Since the hybrid DFT-QMC approach directly borrows the vibrational entropy from the DFT calculation, this explains the high similarity between the DFT and hybrid DFT-QMC values of $\Delta G^\oplus$ and $k_\rightarrow$ for NH$_3$ inversion. In the S$_\textrm{N}$2 system, however, as shown in Table \ref{tab:SN2_thermodynamic_breakdown}, the dominant contribution now becomes the \textit{difference in potential energies} of the transition state and equilibrium state, which DFT predicts to be $\sim$0.008 eV, whereas DMC and CCSD predict to be $\sim$0.022 eV. Since DMC computes the potential energies more accurately than DFT, the hybrid DFT-QMC approach is able to obtain $\Delta G^\oplus$ and $k_\rightarrow$ values in much better agreement with the CCSD result.

\section{Conclusions}
\label{sec:conclusions}
In this work, we have demonstrated the extension of the surrogate Hessian line-search method previously employed to determine equilibrium geometries to the identification of transition states and minimum-energy pathways.
Modifying the objective function of the line-search to maximize the energy along the negative-eigenvalue directions, as shown in Eq. \ref{eq:modified_objective}, permits us to identify saddle points of arbitrary order. Furthermore, performing the line-search within restricted, path-orthogonal subspaces yields points along the MEP that can be specified to arbitrary structural accuracy, leveraging the controllable computational cost of DMC and other stochastic \textit{ab initio} theories. Finally, we have shown that it is possible to combine stochastic potential energies and Monte Carlo-sampled rotational entropies obtained using DMC with vibrational contributions from DFT to compute corrected free energy changes and kinetic constants that can improve upon DFT results and, potentially, approach coupled cluster accuracy using only energetic information from DMC calculations without the need for gradient evaluation at the DMC level.

The surrogate Hessian line-search and subspace optimization methods are both primarily based on energy optimization. Imposing structural tolerances on the geometries that are identified using these methods, therefore, directly influences the precision to which the energy is specified. In practice, the structural tolerance must be set large enough so as not to incur disproportionate computational expense while still meeting the level of precision required for the study. If an accurately designed surrogate Hessian-based calculation results in structures that show slight deviations from more accurate methods such as CCSD or CCSD(T), this is likely an indication of the ``flatness'' of the PES near those structures, which, in turn, reveals interesting information about the physics of the system.

Subspace optimization, as noted in Section \ref{sec:results_mep}, features a few methodological departures from NEB and other chain-of-states methods. However, both classes of methods accomplish the same task and identify slightly different - yet theoretically consistent - points along the MEP. Steps can be taken to eliminate the differences between the two approaches, such as standardizing the tangent calculation method, repeated reevaluation of the surrogate Hessian at every step taken by the line-search, etc. We leave these modifications for future work.

High-accuracy PES calculations play a significant role in several areas of computational chemistry and materials science research. At large system sizes, $\mathcal{O}(N^{6-7})$- or higher-scaling coupled cluster- and full configuration interaction-based methods remain largely inaccessible. Stochastic electronic structure theories such as QMC - accelerated using surrogate theories such as DFT - present opportunities to unravel the physics and chemistry of large, correlated systems at high accuracy and for  potentially lower computational cost, given their $\mathcal{O}(N^{3-4})$ scaling. The work presented here also provides a path toward obtaining critical information about reaction pathways without the explicit use of energy gradients, which can be challenging to obtain in a variety of otherwise promising electronic structure methods.~\cite{Kresse_RPA_Forces} In the context of DMC, our approach will have particular value for large systems or systems containing heavy atoms, for which force calculations can be especially cumbersome.

Overall, our work represents a step further in the expanding suite of capabilities of QMC methods, enabling the accurate description of systems that undergo dynamical transitions along minimum-energy pathways, such as diffusion processes,\cite{Ma_Alfe_Kresse_2011} chemical reactions,\cite{Saccani_JCP_2013} including catalytic reactions,\cite{Shi_JACS_2023,finite_size_error_cancellation} conformational changes, and phase transitions.~\cite{Wines_VSe2}

\section*{Supporting Information}
The Supporting Information document contains the structural parameterization equations for the NH$_3$ and S$_\textrm{N}$2 systems, DMC calculation extension factors, and the breakdown of the free energy contributions. All code associated with this work can be found at
\\
\texttt{https://github.com/gopal-iyer/surrogate\_hessian\_reaction\_pathway\_search}.

\begin{acknowledgement}
G.R.I. thanks Shubham Sharma, Benjamin Foulon, and Bjarne Kreitz for useful discussions. G.R.I. (derivation, implementation, and analysis of the algorithms presented, drafting of the manuscript) was funded by AFOSR Award Number FA9550-19-1-9999 and the Brown University Chemistry Department Vince Wernig Fellowship. J.T. (mentorship) was self-supported.  J.T.K. (concept, mentorship), P.R.C.K. (concept), and B.R. (concept, mentorship, manuscript writing) were supported by the U.S. Department of Energy, Office of Science, Basic Energy Sciences, Materials Sciences and Engineering Division, as part of the Computational Materials Sciences Program and Center for Predictive Simulation of Functional Materials. This research was conducted using computational resources and services at the Center for Computation and Visualization, Brown University.

This manuscript has been authored by UT-Battelle, LLC under Contract No. DE-AC05-00OR22725 with the U.S. Department of Energy. The United States Government retains and the publisher, by accepting the article for publication, acknowledges that the United States Government retains a non-exclusive, paid-up, irrevocable, world-wide license to publish or reproduce the published form of this manuscript, or allow others to do so, for United States Government purposes. 

\end{acknowledgement}



\begin{mcitethebibliography}{64}
\providecommand*\natexlab[1]{#1}
\providecommand*\mciteSetBstSublistMode[1]{}
\providecommand*\mciteSetBstMaxWidthForm[2]{}
\providecommand*\mciteBstWouldAddEndPuncttrue
  {\def\EndOfBibitem{\unskip.}}
\providecommand*\mciteBstWouldAddEndPunctfalse
  {\let\EndOfBibitem\relax}
\providecommand*\mciteSetBstMidEndSepPunct[3]{}
\providecommand*\mciteSetBstSublistLabelBeginEnd[3]{}
\providecommand*\EndOfBibitem{}
\mciteSetBstSublistMode{f}
\mciteSetBstMaxWidthForm{subitem}{(\alph{mcitesubitemcount})}
\mciteSetBstSublistLabelBeginEnd
  {\mcitemaxwidthsubitemform\space}
  {\relax}
  {\relax}

\bibitem[Cramer(2013)]{cramer_computational_chemistry}
Cramer,~C. \emph{Essentials of Computational Chemistry: Theories and Models}; Wiley, 2013\relax
\mciteBstWouldAddEndPuncttrue
\mciteSetBstMidEndSepPunct{\mcitedefaultmidpunct}
{\mcitedefaultendpunct}{\mcitedefaultseppunct}\relax
\EndOfBibitem
\bibitem[Czakó \latin{et~al.}(2014)Czakó, Szabó, and Telekes]{pes_1}
Czakó,~G.; Szabó,~I.; Telekes,~H. On the Choice of the Ab Initio Level of Theory for Potential Energy Surface Developments. \emph{The Journal of Physical Chemistry A} \textbf{2014}, \emph{118}, 646--654\relax
\mciteBstWouldAddEndPuncttrue
\mciteSetBstMidEndSepPunct{\mcitedefaultmidpunct}
{\mcitedefaultendpunct}{\mcitedefaultseppunct}\relax
\EndOfBibitem
\bibitem[Bowman \latin{et~al.}(2011)Bowman, Czak\'o, and Fu]{pes_2}
Bowman,~J.~M.; Czak\'o,~G.; Fu,~B. High-dimensional ab initio potential energy surfaces for reaction dynamics calculations. \emph{Phys. Chem. Chem. Phys.} \textbf{2011}, \emph{13}, 8094--8111\relax
\mciteBstWouldAddEndPuncttrue
\mciteSetBstMidEndSepPunct{\mcitedefaultmidpunct}
{\mcitedefaultendpunct}{\mcitedefaultseppunct}\relax
\EndOfBibitem
\bibitem[Bowman \latin{et~al.}(2011)Bowman, Czakó, and Fu]{high_dimensional_pes}
Bowman,~J.~M.; Czakó,~G.; Fu,~B. High-dimensional ab initio potential energy surfaces for reaction dynamics calculations. \emph{Phys. Chem. Chem. Phys.} \textbf{2011}, \emph{13}, 8094--8111\relax
\mciteBstWouldAddEndPuncttrue
\mciteSetBstMidEndSepPunct{\mcitedefaultmidpunct}
{\mcitedefaultendpunct}{\mcitedefaultseppunct}\relax
\EndOfBibitem
\bibitem[Schlegel(2011)]{geometry_optimization_review}
Schlegel,~H.~B. Geometry optimization. \emph{WIREs Computational Molecular Science} \textbf{2011}, \emph{1}, 790--809\relax
\mciteBstWouldAddEndPuncttrue
\mciteSetBstMidEndSepPunct{\mcitedefaultmidpunct}
{\mcitedefaultendpunct}{\mcitedefaultseppunct}\relax
\EndOfBibitem
\bibitem[Garijo~del R\'{\i}o \latin{et~al.}(2019)Garijo~del R\'{\i}o, Mortensen, and Jacobsen]{gpmin_geometry_optimization}
Garijo~del R\'{\i}o,~E.; Mortensen,~J.~J.; Jacobsen,~K.~W. Local Bayesian optimizer for atomic structures. \emph{Phys. Rev. B} \textbf{2019}, \emph{100}, 104103\relax
\mciteBstWouldAddEndPuncttrue
\mciteSetBstMidEndSepPunct{\mcitedefaultmidpunct}
{\mcitedefaultendpunct}{\mcitedefaultseppunct}\relax
\EndOfBibitem
\bibitem[Bitzek \latin{et~al.}(2006)Bitzek, Koskinen, G\"ahler, Moseler, and Gumbsch]{fire_geometry_optimization}
Bitzek,~E.; Koskinen,~P.; G\"ahler,~F.; Moseler,~M.; Gumbsch,~P. Structural Relaxation Made Simple. \emph{Phys. Rev. Lett.} \textbf{2006}, \emph{97}, 170201\relax
\mciteBstWouldAddEndPuncttrue
\mciteSetBstMidEndSepPunct{\mcitedefaultmidpunct}
{\mcitedefaultendpunct}{\mcitedefaultseppunct}\relax
\EndOfBibitem
\bibitem[Pratt(1986)]{chain_of_states_1}
Pratt,~L.~R. {A statistical method for identifying transition states in high dimensional problems}. \emph{The Journal of Chemical Physics} \textbf{1986}, \emph{85}, 5045--5048\relax
\mciteBstWouldAddEndPuncttrue
\mciteSetBstMidEndSepPunct{\mcitedefaultmidpunct}
{\mcitedefaultendpunct}{\mcitedefaultseppunct}\relax
\EndOfBibitem
\bibitem[Dellago \latin{et~al.}(1998)Dellago, Bolhuis, Csajka, and Chandler]{chain_of_states_2}
Dellago,~C.; Bolhuis,~P.~G.; Csajka,~F.~S.; Chandler,~D. {Transition path sampling and the calculation of rate constants}. \emph{The Journal of Chemical Physics} \textbf{1998}, \emph{108}, 1964--1977\relax
\mciteBstWouldAddEndPuncttrue
\mciteSetBstMidEndSepPunct{\mcitedefaultmidpunct}
{\mcitedefaultendpunct}{\mcitedefaultseppunct}\relax
\EndOfBibitem
\bibitem[JÓNSSON \latin{et~al.}()JÓNSSON, MILLS, and JACOBSEN]{neb_original_paper}
JÓNSSON,~H.; MILLS,~G.; JACOBSEN,~K.~W. \emph{Classical and Quantum Dynamics in Condensed Phase Simulations}; pp 385--404\relax
\mciteBstWouldAddEndPuncttrue
\mciteSetBstMidEndSepPunct{\mcitedefaultmidpunct}
{\mcitedefaultendpunct}{\mcitedefaultseppunct}\relax
\EndOfBibitem
\bibitem[Sheppard \latin{et~al.}(2012)Sheppard, Xiao, Chemelewski, Johnson, and Henkelman]{neb_2}
Sheppard,~D.; Xiao,~P.; Chemelewski,~W.; Johnson,~D.~D.; Henkelman,~G. {A generalized solid-state nudged elastic band method}. \emph{The Journal of Chemical Physics} \textbf{2012}, \emph{136}, 074103\relax
\mciteBstWouldAddEndPuncttrue
\mciteSetBstMidEndSepPunct{\mcitedefaultmidpunct}
{\mcitedefaultendpunct}{\mcitedefaultseppunct}\relax
\EndOfBibitem
\bibitem[Kolsbjerg \latin{et~al.}(2016)Kolsbjerg, Groves, and Hammer]{neb_3}
Kolsbjerg,~E.~L.; Groves,~M.~N.; Hammer,~B. {An automated nudged elastic band method}. \emph{The Journal of Chemical Physics} \textbf{2016}, \emph{145}, 094107\relax
\mciteBstWouldAddEndPuncttrue
\mciteSetBstMidEndSepPunct{\mcitedefaultmidpunct}
{\mcitedefaultendpunct}{\mcitedefaultseppunct}\relax
\EndOfBibitem
\bibitem[Henkelman \latin{et~al.}(2000)Henkelman, Uberuaga, and Jónsson]{ci_neb}
Henkelman,~G.; Uberuaga,~B.~P.; Jónsson,~H. {A climbing image nudged elastic band method for finding saddle points and minimum energy paths}. \emph{The Journal of Chemical Physics} \textbf{2000}, \emph{113}, 9901--9904\relax
\mciteBstWouldAddEndPuncttrue
\mciteSetBstMidEndSepPunct{\mcitedefaultmidpunct}
{\mcitedefaultendpunct}{\mcitedefaultseppunct}\relax
\EndOfBibitem
\bibitem[Alaghemandi \latin{et~al.}(2017)Alaghemandi, Newcomb, and Green]{combustion}
Alaghemandi,~M.; Newcomb,~L.~B.; Green,~J.~R. Ignition in an Atomistic Model of Hydrogen Oxidation. \emph{The Journal of Physical Chemistry A} \textbf{2017}, \emph{121}, 1686--1692, PMID: 28169533\relax
\mciteBstWouldAddEndPuncttrue
\mciteSetBstMidEndSepPunct{\mcitedefaultmidpunct}
{\mcitedefaultendpunct}{\mcitedefaultseppunct}\relax
\EndOfBibitem
\bibitem[Thomas(2009)]{heterogeneous_catalysis_1}
Thomas,~J. Handbook Of Heterogeneous Catalysis. 2., completely revised and enlarged Edition. Vol. 1–8. Edited by G. Ertl, H. Knözinger, F. Schüth, and J. Weitkamp. \emph{Angewandte Chemie International Edition} \textbf{2009}, \emph{48}\relax
\mciteBstWouldAddEndPuncttrue
\mciteSetBstMidEndSepPunct{\mcitedefaultmidpunct}
{\mcitedefaultendpunct}{\mcitedefaultseppunct}\relax
\EndOfBibitem
\bibitem[Nørskov \latin{et~al.}(2014)Nørskov, Studt, Abild-Pedersen, and Bligaard]{heterogeneous_catalysis_2}
Nørskov,~J.~K.; Studt,~F.; Abild-Pedersen,~F.; Bligaard,~T. \emph{Fundamental Concepts in Heterogeneous Catalysis}; John Wiley \& Sons, Inc., 2014; First published: 29 August 2014\relax
\mciteBstWouldAddEndPuncttrue
\mciteSetBstMidEndSepPunct{\mcitedefaultmidpunct}
{\mcitedefaultendpunct}{\mcitedefaultseppunct}\relax
\EndOfBibitem
\bibitem[Wang \latin{et~al.}(2014)Wang, Titov, McGibbon, Liu, Pande, and Mart{\'i}nez]{martinez_nanoreactor}
Wang,~L.-P.; Titov,~A.; McGibbon,~R.; Liu,~F.; Pande,~V.~S.; Mart{\'i}nez,~T.~J. Discovering chemistry with an ab initio nanoreactor. \emph{Nature Chemistry} \textbf{2014}, \emph{6}, 1044--1048\relax
\mciteBstWouldAddEndPuncttrue
\mciteSetBstMidEndSepPunct{\mcitedefaultmidpunct}
{\mcitedefaultendpunct}{\mcitedefaultseppunct}\relax
\EndOfBibitem
\bibitem[Politzer and Murray(2018)Politzer, and Murray]{hellmann_feynman}
Politzer,~P.; Murray,~J.~S. The Hellmann-Feynman theorem: a perspective. \emph{Journal of Molecular Modeling} \textbf{2018}, \emph{24}, 266\relax
\mciteBstWouldAddEndPuncttrue
\mciteSetBstMidEndSepPunct{\mcitedefaultmidpunct}
{\mcitedefaultendpunct}{\mcitedefaultseppunct}\relax
\EndOfBibitem
\bibitem[Salter \latin{et~al.}(1989)Salter, Trucks, and Bartlett]{Salter_JCP_1989}
Salter,~E.~A.; Trucks,~G.~W.; Bartlett,~R.~J. {Analytic energy derivatives in many‐body methods. I. First derivatives}. \emph{The Journal of Chemical Physics} \textbf{1989}, \emph{90}, 1752--1766\relax
\mciteBstWouldAddEndPuncttrue
\mciteSetBstMidEndSepPunct{\mcitedefaultmidpunct}
{\mcitedefaultendpunct}{\mcitedefaultseppunct}\relax
\EndOfBibitem
\bibitem[Stanton and Gauss(2000)Stanton, and Gauss]{Stanton_Gauss_2000}
Stanton,~J.~F.; Gauss,~J. Analytic second derivatives in high-order many-body perturbation and coupled-cluster theories: Computational considerations and applications. \emph{International Reviews in Physical Chemistry} \textbf{2000}, \emph{19}, 61--95\relax
\mciteBstWouldAddEndPuncttrue
\mciteSetBstMidEndSepPunct{\mcitedefaultmidpunct}
{\mcitedefaultendpunct}{\mcitedefaultseppunct}\relax
\EndOfBibitem
\bibitem[Burow \latin{et~al.}(2014)Burow, Bates, Furche, and Eshuis]{burow_analytical_2014}
Burow,~A.~M.; Bates,~J.~E.; Furche,~F.; Eshuis,~H. Analytical {First}-{Order} {Molecular} {Properties} and {Forces} within the {Adiabatic} {Connection} {Random} {Phase} {Approximation}. \emph{Journal of Chemical Theory and Computation} \textbf{2014}, \emph{10}, 180--194, \_eprint: https://doi.org/10.1021/ct4008553\relax
\mciteBstWouldAddEndPuncttrue
\mciteSetBstMidEndSepPunct{\mcitedefaultmidpunct}
{\mcitedefaultendpunct}{\mcitedefaultseppunct}\relax
\EndOfBibitem
\bibitem[Ramberger \latin{et~al.}(2017)Ramberger, Sch\"afer, and Kresse]{Kresse_RPA_Forces}
Ramberger,~B.; Sch\"afer,~T.; Kresse,~G. Analytic Interatomic Forces in the Random Phase Approximation. \emph{Phys. Rev. Lett.} \textbf{2017}, \emph{118}, 106403\relax
\mciteBstWouldAddEndPuncttrue
\mciteSetBstMidEndSepPunct{\mcitedefaultmidpunct}
{\mcitedefaultendpunct}{\mcitedefaultseppunct}\relax
\EndOfBibitem
\bibitem[Assaraf and Caffarel(2000)Assaraf, and Caffarel]{qmc_force_1}
Assaraf,~R.; Caffarel,~M. {Computing forces with quantum Monte Carlo}. \emph{The Journal of Chemical Physics} \textbf{2000}, \emph{113}, 4028--4034\relax
\mciteBstWouldAddEndPuncttrue
\mciteSetBstMidEndSepPunct{\mcitedefaultmidpunct}
{\mcitedefaultendpunct}{\mcitedefaultseppunct}\relax
\EndOfBibitem
\bibitem[Badinski \latin{et~al.}(2010)Badinski, Haynes, Trail, and Needs]{qmc_force_2}
Badinski,~A.; Haynes,~P.~D.; Trail,~J.~R.; Needs,~R.~J. Methods for calculating forces within quantum Monte Carlo simulations. \emph{Journal of Physics: Condensed Matter} \textbf{2010}, \emph{22}, 074202\relax
\mciteBstWouldAddEndPuncttrue
\mciteSetBstMidEndSepPunct{\mcitedefaultmidpunct}
{\mcitedefaultendpunct}{\mcitedefaultseppunct}\relax
\EndOfBibitem
\bibitem[Chiesa \latin{et~al.}(2005)Chiesa, Ceperley, and Zhang]{qmc_force_3}
Chiesa,~S.; Ceperley,~D.~M.; Zhang,~S. Accurate, Efficient, and Simple Forces Computed with Quantum Monte Carlo Methods. \emph{Phys. Rev. Lett.} \textbf{2005}, \emph{94}, 036404\relax
\mciteBstWouldAddEndPuncttrue
\mciteSetBstMidEndSepPunct{\mcitedefaultmidpunct}
{\mcitedefaultendpunct}{\mcitedefaultseppunct}\relax
\EndOfBibitem
\bibitem[Filippi and Umrigar(2000)Filippi, and Umrigar]{Filippi_PRB_2000}
Filippi,~C.; Umrigar,~C.~J. Correlated sampling in quantum Monte Carlo: A route to forces. \emph{Phys. Rev. B} \textbf{2000}, \emph{61}, R16291--R16294\relax
\mciteBstWouldAddEndPuncttrue
\mciteSetBstMidEndSepPunct{\mcitedefaultmidpunct}
{\mcitedefaultendpunct}{\mcitedefaultseppunct}\relax
\EndOfBibitem
\bibitem[Assaraf and Caffarel(2003)Assaraf, and Caffarel]{Assaraf_Cafferel_2003}
Assaraf,~R.; Caffarel,~M. {Zero-variance zero-bias principle for observables in quantum Monte Carlo: Application to forces}. \emph{The Journal of Chemical Physics} \textbf{2003}, \emph{119}, 10536--10552\relax
\mciteBstWouldAddEndPuncttrue
\mciteSetBstMidEndSepPunct{\mcitedefaultmidpunct}
{\mcitedefaultendpunct}{\mcitedefaultseppunct}\relax
\EndOfBibitem
\bibitem[R\'{\i}os and Conduit(2019)R\'{\i}os, and Conduit]{Rios_PRE_2019}
R\'{\i}os,~P.~L.; Conduit,~G.~J. Tail-regression estimator for heavy-tailed distributions of known tail indices and its application to continuum quantum Monte Carlo data. \emph{Phys. Rev. E} \textbf{2019}, \emph{99}, 063312\relax
\mciteBstWouldAddEndPuncttrue
\mciteSetBstMidEndSepPunct{\mcitedefaultmidpunct}
{\mcitedefaultendpunct}{\mcitedefaultseppunct}\relax
\EndOfBibitem
\bibitem[Nakano \latin{et~al.}(2022)Nakano, Raghav, and Sorella]{Nakano_Sorella_2022}
Nakano,~K.; Raghav,~A.; Sorella,~S. {Space-warp coordinate transformation for efficient ionic force calculations in quantum Monte Carlo}. \emph{The Journal of Chemical Physics} \textbf{2022}, \emph{156}, 034101\relax
\mciteBstWouldAddEndPuncttrue
\mciteSetBstMidEndSepPunct{\mcitedefaultmidpunct}
{\mcitedefaultendpunct}{\mcitedefaultseppunct}\relax
\EndOfBibitem
\bibitem[Moroni \latin{et~al.}(2014)Moroni, Saccani, and Filippi]{moroni_practical_2014}
Moroni,~S.; Saccani,~S.; Filippi,~C. Practical {Schemes} for {Accurate} {Forces} in {Quantum} {Monte} {Carlo}. \emph{Journal of Chemical Theory and Computation} \textbf{2014}, \emph{10}, 4823--4829, \_eprint: https://doi.org/10.1021/ct500780r\relax
\mciteBstWouldAddEndPuncttrue
\mciteSetBstMidEndSepPunct{\mcitedefaultmidpunct}
{\mcitedefaultendpunct}{\mcitedefaultseppunct}\relax
\EndOfBibitem
\bibitem[Jiang \latin{et~al.}(2022)Jiang, Fang, Alavi, and Chen]{qmc_force_4}
Jiang,~T.; Fang,~W.; Alavi,~A.; Chen,~J. General Analytical Nuclear Forces and Molecular Potential Energy Surface from Full Configuration Interaction Quantum Monte Carlo. \emph{Journal of Chemical Theory and Computation} \textbf{2022}, \emph{18}, 7233--7242, PMID: 36326847\relax
\mciteBstWouldAddEndPuncttrue
\mciteSetBstMidEndSepPunct{\mcitedefaultmidpunct}
{\mcitedefaultendpunct}{\mcitedefaultseppunct}\relax
\EndOfBibitem
\bibitem[Chen and Zhang(2022)Chen, and Zhang]{chen_structural_2022}
Chen,~S.; Zhang,~S. A structural optimization algorithm with stochastic forces and stresses. \emph{Nature Computational Science} \textbf{2022}, \emph{2}, 736--744\relax
\mciteBstWouldAddEndPuncttrue
\mciteSetBstMidEndSepPunct{\mcitedefaultmidpunct}
{\mcitedefaultendpunct}{\mcitedefaultseppunct}\relax
\EndOfBibitem
\bibitem[Iyer and Rubenstein(2022)Iyer, and Rubenstein]{finite_size_error_cancellation}
Iyer,~G.~R.; Rubenstein,~B.~M. Finite-Size Error Cancellation in Diffusion Monte Carlo Calculations of Surface Chemistry. \emph{The Journal of Physical Chemistry A} \textbf{2022}, \emph{126}, 4636--4646\relax
\mciteBstWouldAddEndPuncttrue
\mciteSetBstMidEndSepPunct{\mcitedefaultmidpunct}
{\mcitedefaultendpunct}{\mcitedefaultseppunct}\relax
\EndOfBibitem
\bibitem[Staros \latin{et~al.}(2022)Staros, Hu, Tiihonen, Nanguneri, Krogel, Bennett, Heinonen, Ganesh, and Rubenstein]{CrI3_dmc}
Staros,~D.; Hu,~G.; Tiihonen,~J.; Nanguneri,~R.; Krogel,~J.; Bennett,~M.~C.; Heinonen,~O.; Ganesh,~P.; Rubenstein,~B. {A combined first principles study of the structural, magnetic, and phonon properties of monolayer CrI3}. \emph{The Journal of Chemical Physics} \textbf{2022}, \emph{156}, 014707\relax
\mciteBstWouldAddEndPuncttrue
\mciteSetBstMidEndSepPunct{\mcitedefaultmidpunct}
{\mcitedefaultendpunct}{\mcitedefaultseppunct}\relax
\EndOfBibitem
\bibitem[Shin \latin{et~al.}(2021)Shin, Krogel, Gasperich, Kent, Benali, and Heinonen]{GeSe_dmc}
Shin,~H.; Krogel,~J.~T.; Gasperich,~K.; Kent,~P. R.~C.; Benali,~A.; Heinonen,~O. Optimized structure and electronic band gap of monolayer GeSe from quantum Monte Carlo methods. \emph{Phys. Rev. Mater.} \textbf{2021}, \emph{5}, 024002\relax
\mciteBstWouldAddEndPuncttrue
\mciteSetBstMidEndSepPunct{\mcitedefaultmidpunct}
{\mcitedefaultendpunct}{\mcitedefaultseppunct}\relax
\EndOfBibitem
\bibitem[Ryczko \latin{et~al.}(2022)Ryczko, Krogel, and Tamblyn]{qmc_ml_energies}
Ryczko,~K.; Krogel,~J.~T.; Tamblyn,~I. Machine Learning Diffusion Monte Carlo Energies. \emph{Journal of Chemical Theory and Computation} \textbf{2022}, \emph{18}, 7695--7701, PMID: 36317712\relax
\mciteBstWouldAddEndPuncttrue
\mciteSetBstMidEndSepPunct{\mcitedefaultmidpunct}
{\mcitedefaultendpunct}{\mcitedefaultseppunct}\relax
\EndOfBibitem
\bibitem[Huang and Rubenstein(2023)Huang, and Rubenstein]{qmc_force_5}
Huang,~C.; Rubenstein,~B.~M. Machine Learning Diffusion Monte Carlo Forces. \emph{The Journal of Physical Chemistry A} \textbf{2023}, \emph{127}, 339--355, PMID: 36576803\relax
\mciteBstWouldAddEndPuncttrue
\mciteSetBstMidEndSepPunct{\mcitedefaultmidpunct}
{\mcitedefaultendpunct}{\mcitedefaultseppunct}\relax
\EndOfBibitem
\bibitem[Chen \latin{et~al.}(2023)Chen, Lee, Ye, Berkelbach, Reichman, and Markland]{transfer_learning_potential}
Chen,~M.~S.; Lee,~J.; Ye,~H.-Z.; Berkelbach,~T.~C.; Reichman,~D.~R.; Markland,~T.~E. Data-Efficient Machine Learning Potentials from Transfer Learning of Periodic Correlated Electronic Structure Methods: Liquid Water at AFQMC, CCSD, and CCSD(T) Accuracy. \emph{Journal of Chemical Theory and Computation} \textbf{2023}, \emph{19}, 4510--4519, PMID: 36730728\relax
\mciteBstWouldAddEndPuncttrue
\mciteSetBstMidEndSepPunct{\mcitedefaultmidpunct}
{\mcitedefaultendpunct}{\mcitedefaultseppunct}\relax
\EndOfBibitem
\bibitem[Archibald \latin{et~al.}(2018)Archibald, Krogel, and Kent]{Archibald_JCP_2018}
Archibald,~R.; Krogel,~J.~T.; Kent,~P. R.~C. {Gaussian process based optimization of molecular geometries using statistically sampled energy surfaces from quantum Monte Carlo}. \emph{The Journal of Chemical Physics} \textbf{2018}, \emph{149}, 164116\relax
\mciteBstWouldAddEndPuncttrue
\mciteSetBstMidEndSepPunct{\mcitedefaultmidpunct}
{\mcitedefaultendpunct}{\mcitedefaultseppunct}\relax
\EndOfBibitem
\bibitem[Ceperley \latin{et~al.}(2023)Ceperley, Jensen, Yang, Niu, Pierleoni, and Holzmann]{ceperley2023training}
Ceperley,~D.~M.; Jensen,~S.; Yang,~Y.; Niu,~H.; Pierleoni,~C.; Holzmann,~M. Training models using forces computed by stochastic electronic structure methods. 2023\relax
\mciteBstWouldAddEndPuncttrue
\mciteSetBstMidEndSepPunct{\mcitedefaultmidpunct}
{\mcitedefaultendpunct}{\mcitedefaultseppunct}\relax
\EndOfBibitem
\bibitem[Grossman and Mitas(2005)Grossman, and Mitas]{Grossman_PRL}
Grossman,~J.~C.; Mitas,~L. Efficient Quantum Monte Carlo Energies for Molecular Dynamics Simulations. \emph{Phys. Rev. Lett.} \textbf{2005}, \emph{94}, 056403\relax
\mciteBstWouldAddEndPuncttrue
\mciteSetBstMidEndSepPunct{\mcitedefaultmidpunct}
{\mcitedefaultendpunct}{\mcitedefaultseppunct}\relax
\EndOfBibitem
\bibitem[Tiihonen \latin{et~al.}(2022)Tiihonen, Kent, and Krogel]{original_paper}
Tiihonen,~J.; Kent,~P. R.~C.; Krogel,~J.~T. {Surrogate Hessian accelerated structural optimization for stochastic electronic structure theories}. \emph{The Journal of Chemical Physics} \textbf{2022}, \emph{156}, 054104\relax
\mciteBstWouldAddEndPuncttrue
\mciteSetBstMidEndSepPunct{\mcitedefaultmidpunct}
{\mcitedefaultendpunct}{\mcitedefaultseppunct}\relax
\EndOfBibitem
\bibitem[Samanta and E(2013)Samanta, and E]{optimization_string_method}
Samanta,~A.; E,~W. Optimization-Based String Method for Finding Minimum Energy Path. \emph{Communications in Computational Physics} \textbf{2013}, \emph{14}, 265–275\relax
\mciteBstWouldAddEndPuncttrue
\mciteSetBstMidEndSepPunct{\mcitedefaultmidpunct}
{\mcitedefaultendpunct}{\mcitedefaultseppunct}\relax
\EndOfBibitem
\bibitem[Saccani \latin{et~al.}(2013)Saccani, Filippi, and Moroni]{Saccani_JCP_2013}
Saccani,~S.; Filippi,~C.; Moroni,~S. {Minimum energy pathways via quantum Monte Carlo}. \emph{The Journal of Chemical Physics} \textbf{2013}, \emph{138}, 084109\relax
\mciteBstWouldAddEndPuncttrue
\mciteSetBstMidEndSepPunct{\mcitedefaultmidpunct}
{\mcitedefaultendpunct}{\mcitedefaultseppunct}\relax
\EndOfBibitem
\bibitem[Gonzales \latin{et~al.}(2001)Gonzales, Cox, Brown, Allen, and Schaefer]{reactions}
Gonzales,~J.~M.; Cox,~R.~S.; Brown,~S.~T.; Allen,~W.~D.; Schaefer,~H.~F. Assessment of Density Functional Theory for Model SN2 Reactions: CH3X + F- (X = F, Cl, CN, OH, SH, NH2, PH2). \emph{The Journal of Physical Chemistry A} \textbf{2001}, \emph{105}, 11327--11346\relax
\mciteBstWouldAddEndPuncttrue
\mciteSetBstMidEndSepPunct{\mcitedefaultmidpunct}
{\mcitedefaultendpunct}{\mcitedefaultseppunct}\relax
\EndOfBibitem
\bibitem[Foulkes \latin{et~al.}(2001)Foulkes, Mitas, Needs, and Rajagopal]{foulkes_qmc}
Foulkes,~W. M.~C.; Mitas,~L.; Needs,~R.~J.; Rajagopal,~G. Quantum Monte Carlo simulations of solids. \emph{Rev. Mod. Phys.} \textbf{2001}, \emph{73}, 33--83\relax
\mciteBstWouldAddEndPuncttrue
\mciteSetBstMidEndSepPunct{\mcitedefaultmidpunct}
{\mcitedefaultendpunct}{\mcitedefaultseppunct}\relax
\EndOfBibitem
\bibitem[Ceperley(2010)]{ceperley_qmc}
Ceperley,~D.~M. In \emph{Theoretical and Computational Methods in Mineral Physics}; Wentzcovitch,~R.~M., Stixrude,~L., Eds.; De Gruyter: Berlin, Boston, 2010; pp 129--136\relax
\mciteBstWouldAddEndPuncttrue
\mciteSetBstMidEndSepPunct{\mcitedefaultmidpunct}
{\mcitedefaultendpunct}{\mcitedefaultseppunct}\relax
\EndOfBibitem
\bibitem[Fukui(1970)]{fukui_irc}
Fukui,~K. Formulation of the reaction coordinate. \emph{The Journal of Physical Chemistry} \textbf{1970}, \emph{74}, 4161--4163\relax
\mciteBstWouldAddEndPuncttrue
\mciteSetBstMidEndSepPunct{\mcitedefaultmidpunct}
{\mcitedefaultendpunct}{\mcitedefaultseppunct}\relax
\EndOfBibitem
\bibitem[Motagamwala and Dumesic(2021)Motagamwala, and Dumesic]{microkinetic_modeling}
Motagamwala,~A.~H.; Dumesic,~J.~A. Microkinetic Modeling: A Tool for Rational Catalyst Design. \emph{Chemical Reviews} \textbf{2021}, \emph{121}, 1049--1076\relax
\mciteBstWouldAddEndPuncttrue
\mciteSetBstMidEndSepPunct{\mcitedefaultmidpunct}
{\mcitedefaultendpunct}{\mcitedefaultseppunct}\relax
\EndOfBibitem
\bibitem[Anderson(1972)]{more_is_different}
Anderson,~P.~W. More Is Different. \emph{Science} \textbf{1972}, \emph{177}, 393--396\relax
\mciteBstWouldAddEndPuncttrue
\mciteSetBstMidEndSepPunct{\mcitedefaultmidpunct}
{\mcitedefaultendpunct}{\mcitedefaultseppunct}\relax
\EndOfBibitem
\bibitem[Sun \latin{et~al.}(2018)Sun, Berkelbach, Blunt, Booth, Guo, Li, Liu, McClain, Sayfutyarova, Sharma, Wouters, and Chan]{pyscf_1}
Sun,~Q.; Berkelbach,~T.~C.; Blunt,~N.~S.; Booth,~G.~H.; Guo,~S.; Li,~Z.; Liu,~J.; McClain,~J.~D.; Sayfutyarova,~E.~R.; Sharma,~S.; Wouters,~S.; Chan,~G. K.-L. PySCF: the Python-based simulations of chemistry framework. \emph{WIREs Computational Molecular Science} \textbf{2018}, \emph{8}, e1340\relax
\mciteBstWouldAddEndPuncttrue
\mciteSetBstMidEndSepPunct{\mcitedefaultmidpunct}
{\mcitedefaultendpunct}{\mcitedefaultseppunct}\relax
\EndOfBibitem
\bibitem[Sun \latin{et~al.}(2020)Sun, Zhang, Banerjee, Bao, Barbry, Blunt, Bogdanov, Booth, Chen, Cui, Eriksen, Gao, Guo, Hermann, Hermes, Koh, Koval, Lehtola, Li, Liu, Mardirossian, McClain, Motta, Mussard, Pham, Pulkin, Purwanto, Robinson, Ronca, Sayfutyarova, Scheurer, Schurkus, Smith, Sun, Sun, Upadhyay, Wagner, Wang, White, Whitfield, Williamson, Wouters, Yang, Yu, Zhu, Berkelbach, Sharma, Sokolov, and Chan]{pyscf_2}
Sun,~Q. \latin{et~al.}  {Recent developments in the PySCF program package}. \emph{The Journal of Chemical Physics} \textbf{2020}, \emph{153}, 024109\relax
\mciteBstWouldAddEndPuncttrue
\mciteSetBstMidEndSepPunct{\mcitedefaultmidpunct}
{\mcitedefaultendpunct}{\mcitedefaultseppunct}\relax
\EndOfBibitem
\bibitem[Annaberdiyev \latin{et~al.}(2018)Annaberdiyev, Wang, Melton, Bennett, Shulenburger, and Mitas]{ccecp}
Annaberdiyev,~A.; Wang,~G.; Melton,~C.~A.; Bennett,~M.~C.; Shulenburger,~L.; Mitas,~L. {A new generation of effective core potentials from correlated calculations: 3d transition metal series}. \emph{The Journal of Chemical Physics} \textbf{2018}, \emph{149}, 134108\relax
\mciteBstWouldAddEndPuncttrue
\mciteSetBstMidEndSepPunct{\mcitedefaultmidpunct}
{\mcitedefaultendpunct}{\mcitedefaultseppunct}\relax
\EndOfBibitem
\bibitem[Wang \latin{et~al.}(2019)Wang, Annaberdiyev, Melton, Bennett, Shulenburger, and Mitas]{ccecp_2}
Wang,~G.; Annaberdiyev,~A.; Melton,~C.~A.; Bennett,~M.~C.; Shulenburger,~L.; Mitas,~L. {A new generation of effective core potentials from correlated calculations: 4s and 4p main group elements and first row additions}. \emph{The Journal of Chemical Physics} \textbf{2019}, \emph{151}, 144110\relax
\mciteBstWouldAddEndPuncttrue
\mciteSetBstMidEndSepPunct{\mcitedefaultmidpunct}
{\mcitedefaultendpunct}{\mcitedefaultseppunct}\relax
\EndOfBibitem
\bibitem[Perdew \latin{et~al.}(1996)Perdew, Burke, and Ernzerhof]{pbe_functional}
Perdew,~J.~P.; Burke,~K.; Ernzerhof,~M. Generalized Gradient Approximation Made Simple. \emph{Phys. Rev. Lett.} \textbf{1996}, \emph{77}, 3865--3868\relax
\mciteBstWouldAddEndPuncttrue
\mciteSetBstMidEndSepPunct{\mcitedefaultmidpunct}
{\mcitedefaultendpunct}{\mcitedefaultseppunct}\relax
\EndOfBibitem
\bibitem[Kim \latin{et~al.}(2018)Kim, Baczewski, Beaudet, Benali, Bennett, Berrill, Blunt, Borda, Casula, Ceperley, Chiesa, Clark, Clay, Delaney, Dewing, Esler, Hao, Heinonen, Kent, Krogel, Kylänpää, Li, Lopez, Luo, Malone, Martin, Mathuriya, McMinis, Melton, Mitas, Morales, Neuscamman, Parker, Flores, Romero, Rubenstein, Shea, Shin, Shulenburger, Tillack, Townsend, Tubman, Goetz, Vincent, Yang, Yang, Zhang, and Zhao]{qmcpack_1}
Kim,~J. \latin{et~al.}  QMCPACK: an open source ab initio quantum Monte Carlo package for the electronic structure of atoms, molecules and solids. \emph{Journal of Physics: Condensed Matter} \textbf{2018}, \emph{30}, 195901\relax
\mciteBstWouldAddEndPuncttrue
\mciteSetBstMidEndSepPunct{\mcitedefaultmidpunct}
{\mcitedefaultendpunct}{\mcitedefaultseppunct}\relax
\EndOfBibitem
\bibitem[Kent \latin{et~al.}(2020)Kent, Annaberdiyev, Benali, Bennett, Landinez~Borda, Doak, Hao, Jordan, Krogel, Kylänpää, Lee, Luo, Malone, Melton, Mitas, Morales, Neuscamman, Reboredo, Rubenstein, Saritas, Upadhyay, Wang, Zhang, and Zhao]{qmcpack_2}
Kent,~P. R.~C. \latin{et~al.}  {QMCPACK: Advances in the development, efficiency, and application of auxiliary field and real-space variational and diffusion quantum Monte Carlo}. \emph{The Journal of Chemical Physics} \textbf{2020}, \emph{152}, 174105\relax
\mciteBstWouldAddEndPuncttrue
\mciteSetBstMidEndSepPunct{\mcitedefaultmidpunct}
{\mcitedefaultendpunct}{\mcitedefaultseppunct}\relax
\EndOfBibitem
\bibitem[Krogel(2016)]{nexus}
Krogel,~J.~T. Nexus: A modular workflow management system for quantum simulation codes. \emph{Computer Physics Communications} \textbf{2016}, \emph{198}, 154--168\relax
\mciteBstWouldAddEndPuncttrue
\mciteSetBstMidEndSepPunct{\mcitedefaultmidpunct}
{\mcitedefaultendpunct}{\mcitedefaultseppunct}\relax
\EndOfBibitem
\bibitem[Larsen \latin{et~al.}(2017)Larsen, Mortensen, Blomqvist, Castelli, Christensen, Dułak, Friis, Groves, Hammer, Hargus, Hermes, Jennings, Jensen, Kermode, Kitchin, Kolsbjerg, Kubal, Kaasbjerg, Lysgaard, Maronsson, Maxson, Olsen, Pastewka, Peterson, Rostgaard, Schiøtz, Schütt, Strange, Thygesen, Vegge, Vilhelmsen, Walter, Zeng, and Jacobsen]{ase}
Larsen,~A.~H. \latin{et~al.}  The atomic simulation environment—a Python library for working with atoms. \emph{Journal of Physics: Condensed Matter} \textbf{2017}, \emph{29}, 273002\relax
\mciteBstWouldAddEndPuncttrue
\mciteSetBstMidEndSepPunct{\mcitedefaultmidpunct}
{\mcitedefaultendpunct}{\mcitedefaultseppunct}\relax
\EndOfBibitem
\bibitem[Zhu \latin{et~al.}(2019)Zhu, Thompson, and Martínez]{geodesic_interpolation}
Zhu,~X.; Thompson,~K.~C.; Martínez,~T.~J. {Geodesic interpolation for reaction pathways}. \emph{The Journal of Chemical Physics} \textbf{2019}, \emph{150}, 164103\relax
\mciteBstWouldAddEndPuncttrue
\mciteSetBstMidEndSepPunct{\mcitedefaultmidpunct}
{\mcitedefaultendpunct}{\mcitedefaultseppunct}\relax
\EndOfBibitem
\bibitem[Ma \latin{et~al.}(2011)Ma, Michaelides, Alf\`e, Schimka, Kresse, and Wang]{Ma_Alfe_Kresse_2011}
Ma,~J.; Michaelides,~A.; Alf\`e,~D.; Schimka,~L.; Kresse,~G.; Wang,~E. Adsorption and diffusion of water on graphene from first principles. \emph{Phys. Rev. B} \textbf{2011}, \emph{84}, 033402\relax
\mciteBstWouldAddEndPuncttrue
\mciteSetBstMidEndSepPunct{\mcitedefaultmidpunct}
{\mcitedefaultendpunct}{\mcitedefaultseppunct}\relax
\EndOfBibitem
\bibitem[Shi \latin{et~al.}(2023)Shi, Zen, Kapil, Nagy, Grüneis, and Michaelides]{Shi_JACS_2023}
Shi,~B.~X.; Zen,~A.; Kapil,~V.; Nagy,~P.~R.; Grüneis,~A.; Michaelides,~A. Many-Body Methods for Surface Chemistry Come of Age: Achieving Consensus with Experiments. \emph{Journal of the American Chemical Society} \textbf{2023}, \emph{145}, 25372--25381, PMID: 37948071\relax
\mciteBstWouldAddEndPuncttrue
\mciteSetBstMidEndSepPunct{\mcitedefaultmidpunct}
{\mcitedefaultendpunct}{\mcitedefaultseppunct}\relax
\EndOfBibitem
\bibitem[Wines \latin{et~al.}(2023)Wines, Tiihonen, Saritas, Krogel, and Ataca]{Wines_VSe2}
Wines,~D.; Tiihonen,~J.; Saritas,~K.; Krogel,~J.~T.; Ataca,~C. A Quantum Monte Carlo Study of the Structural, Energetic, and Magnetic Properties of Two-Dimensional H and T Phase VSe2. \emph{The Journal of Physical Chemistry Letters} \textbf{2023}, \emph{14}, 3553--3560, PMID: 37017431\relax
\mciteBstWouldAddEndPuncttrue
\mciteSetBstMidEndSepPunct{\mcitedefaultmidpunct}
{\mcitedefaultendpunct}{\mcitedefaultseppunct}\relax
\EndOfBibitem
\end{mcitethebibliography}

\providecommand{\latin}[1]{#1}
\makeatletter
\providecommand{\doi}
  {\begingroup\let\do\@makeother\dospecials
  \catcode`\{=1 \catcode`\}=2 \doi@aux}
\providecommand{\doi@aux}[1]{\endgroup\texttt{#1}}
\makeatother
\providecommand*\mcitethebibliography{\thebibliography}
\csname @ifundefined\endcsname{endmcitethebibliography}  {\let\endmcitethebibliography\endthebibliography}{}


\end{document}


\section{Mapping Structural Parameters to Cartesian Coordinates}
\label{supporting:mapping}
For the NH$_3$ inversion system, we can map the structural parameters to Cartesian coordinates without loss of generality as follows:
\begin{align}
    r_\textrm{N} &= (0,0,0), \\
    r_{\textrm{H}_1} &= \left(-p_0 \cos p_1 , p_0\sin p_1, 0\right),\\
    r_{\textrm{H}_{2x}} &= -p_0 \cos p_1,\\
    r_{\textrm{H}_{2y}} &= -p_0\sin p_1 \sin\left(\frac{\pi}{6}\right),\\
    r_{\textrm{H}_{2z}} &= p_0\sin p_1 \cos\left(\frac{\pi}{6}\right),\\
    r_{\textrm{H}_{3x}} &= -p_0 \cos p_1 ,\\
    r_{\textrm{H}_{3y}} &= -p_0\sin p_1 \sin\left(\frac{\pi}{6}\right),\\
    r_{\textrm{H}_{3z}} &= -p_0\sin p_1 \cos\left(\frac{\pi}{6}\right).
\end{align}

The parameterization of the S$_\textrm{N}$2 system is largely similar, with the addition of two parameters ($p_2$ and $p_3$) to describe the two C-F bonds:
\begin{align}
    r_\textrm{C} &= (0,0,0), \\
    r_{\textrm{H}_1} &= \left(-p_0 \cos p_1 , p_0\sin p_1, 0\right),\\
    r_{\textrm{H}_{2x}} &= -p_0 \cos p_1,\\
    r_{\textrm{H}_{2y}} &= -p_0\sin p_1 \sin\left(\frac{\pi}{6}\right),\\
    r_{\textrm{H}_{2z}} &= p_0\sin p_1 \cos\left(\frac{\pi}{6}\right),\\
    r_{\textrm{H}_{3x}} &= -p_0 \cos p_1 ,\\
    r_{\textrm{H}_{3y}} &= -p_0\sin p_1 \sin\left(\frac{\pi}{6}\right),\\
    r_{\textrm{H}_{3z}} &= -p_0\sin p_1 \cos\left(\frac{\pi}{6}\right),\\
    r_{\textrm{F}_1} &= (-p_2, 0, 0),\\
    r_{\textrm{F}_2} &= (p_3, 0, 0).
\end{align}

\section{Extension of DMC Runs}
\label{supporting:extension}
In order to obtain non-overlapping error bars and precise structural parameters, we run our DMC calculations for a larger number of steps than the initial estimate based upon line search statistical tolerances, as discussed in Section \ref{sec:computational_details}. Specifically, if an estimate requires $N$ DMC steps, we perform the DMC calculation for $(N\times f)$ steps, where $f$ is termed the DMC extension factor. We report the DMC extension factors for NH$_3$ inversion in Table \ref{tab:NH3_DMC_extension_factors} and the S$_\textrm{N}$2 reaction in Table \ref{tab:SN2_DMC_extension_factors}.

\begin{table}
\centering
\begin{tabular}{|c|c|c|c|c|c|}
\hline
     & \textbf{Image 0 (initial)} & \textbf{Image 1} & \textbf{Image 2} & \textbf{Image 3} & \textbf{Image 4 (final)} \\ \hline
$f$ & 64 & 16 & 16 & 64 & 16 \\ \hline
\end{tabular}
\caption{DMC extension factors for the NH$_3$ inversion system. Here, each ``image" is a structure along the MEP.}
\label{tab:NH3_DMC_extension_factors}
\end{table}

\begin{table}
\centering
\begin{tabular}{|c|c|c|c|c|c|c|c|}
\hline
     & \textbf{Image 0 (initial)} & \textbf{Image 1} & \textbf{Image 2} & \textbf{Image 3} & \textbf{Image 4} & \textbf{Image 5} & \textbf{Image 6 (final)} \\ \hline
$f$ & 16 & 16 & 16 & 16 & 16 & 16 & 16 \\ \hline
\end{tabular}
\caption{DMC extension factors for the S$_\textrm{N}$2 reaction. Here, each ``image" is a structure along the MEP.}
\label{tab:SN2_DMC_extension_factors}
\end{table}

\section{Breakdown of Contributions to Free Energies}
Tables \ref{tab:NH3_thermodynamic_breakdown} and \ref{tab:SN2_thermodynamic_breakdown} report the various energetic contributions to the free energy computed using DFT, CCSD, and the hybrid DFT-QMC approach discussed in Sections \ref{sec:hybrid_dft_qmc} and \ref{sec:results_hybrid_dft_qmc}. For clarity, we only report the quantities that depend on the method (DFT, DMC, or CCSD) used to compute them. These are the potential energy ($E_\textrm{pot}$), the zero-point and vibrational contribution to the enthalpy ($E_\textrm{ZPE}^\textrm{DFT} + \int_0^T C_{V,\textrm{vib}}dT$), the rotational entropy ($S_\textrm{rot}$), and the vibrational entropy ($S_\textrm{vib}$).

\begin{table}
\centering
\begin{tabular}{|c|c|c|c|c|}
\hline
    \multirow{2}{*}{} & \multicolumn{4}{|c|}{\textbf{Equilibrium A}} \\ \cline{2-5}
    & $E_\textrm{pot}$ & $E_\textrm{ZPE}^\textrm{DFT} + \int_0^T C_{V,\textrm{vib}}dT$ & $T\times S_\textrm{rot}$ & $T\times S_\textrm{vib}$ \\ \hline
    \textbf{DFT} & -11.726 & 0.192 & 0.148 & \textbf{0.112} \\ \hline
    \textbf{Hybrid DFT-DMC} & -11.729(0) & (same as DFT) & 0.147(0) & (same as DFT) \\ \hline
    \textbf{CCSD} & -11.708 & 0.196 & 0.148 & \textbf{0.107} \\ \hline
\hline
    \multirow{2}{*}{} & \multicolumn{4}{|c|}{\textbf{Transition State}} \\ \cline{2-5}
    & $E_\textrm{pot}$ & $E_\textrm{ZPE}^\textrm{DFT} + \int_0^T C_{V,\textrm{vib}}dT$ & $T\times S_\textrm{rot}$ & $T\times S_\textrm{vib}$ \\ \hline
    \textbf{DFT} & -11.719 & 0.196 & 0.146 & \textbf{0.325} \\ \hline
    \textbf{Hybrid DFT-DMC} & -11.721(0) & (same as DFT) & 0.145(0) & (same as DFT) \\ \hline
    \textbf{CCSD} & -11.699 & 0.198 & 0.146 & \textbf{0.342} \\ \hline
\hline
    \multirow{2}{*}{} & \multicolumn{4}{|c|}{\textbf{Equilibrium B}} \\ \cline{2-5}
    & $E_\textrm{pot}$ & $E_\textrm{ZPE}^\textrm{DFT} + \int_0^T C_{V,\textrm{vib}}dT$ & $T\times S_\textrm{rot}$ & $T\times S_\textrm{vib}$ \\ \hline
    \textbf{DFT} & -11.726 & 0.192 & 0.148 & \textbf{0.112} \\ \hline
    \textbf{Hybrid DFT-DMC} & -11.730(1) & (same as DFT) & 0.145(0) & (same as DFT) \\ \hline
    \textbf{CCSD} & -11.708 & 0.196 & 0.148 & \textbf{0.107} \\ \hline
\end{tabular}
\caption{Breakdown of contributions to free energy for the equilibrium states and transition state of the NH$_3$ system. The dominant contribution (highlighted in bold) to $\Delta G^\oplus$ is the difference in vibrational entropy between the transition state and the equilibrium state. Since this quantity is borrowed directly from DFT in the hybrid DFT-QMC approach, the free energy difference between the equilibrium and transition states in the hybrid approach largely matches the DFT result. All energies are in eV and $T=298$ K.}
\label{tab:NH3_thermodynamic_breakdown}
\end{table}

\begin{table}
\centering
\begin{tabular}{|c|c|c|c|c|}
\hline
    \multirow{2}{*}{} & \multicolumn{4}{|c|}{\textbf{Equilibrium A}} \\ \cline{2-5}
    & $E_\textrm{pot}$ & $E_\textrm{ZPE}^\textrm{DFT} + \int_0^T C_{V,\textrm{vib}}dT$ & $T\times S_\textrm{rot}$ & $T\times S_\textrm{vib}$ \\ \hline
    \textbf{DFT} & \textbf{-56.095} & 0.342 & 0.267 & 0.511 \\ \hline
    \textbf{Hybrid DFT-DMC} & \textbf{-56.093(0)} & (same as DFT) & 0.267(0) & (same as DFT) \\ \hline
    \textbf{CCSD} & \textbf{-55.985} & 0.345 & 0.266 & 0.477 \\ \hline
\hline
    \multirow{2}{*}{} & \multicolumn{4}{|c|}{\textbf{Transition State}} \\ \cline{2-5}
    & $E_\textrm{pot}$ & $E_\textrm{ZPE}^\textrm{DFT} + \int_0^T C_{V,\textrm{vib}}dT$ & $T\times S_\textrm{rot}$ & $T\times S_\textrm{vib}$ \\ \hline
    \textbf{DFT} & \textbf{-56.087} & 0.344 & 0.264 & 0.520 \\ \hline
    \textbf{Hybrid DFT-DMC} & \textbf{-56.070(1)} & (same as DFT) & 0.262(0) & (same as DFT) \\ \hline
    \textbf{CCSD} & \textbf{-55.963} & 0.347 & 0.263 & 0.484 \\ \hline
\hline
    \multirow{2}{*}{} & \multicolumn{4}{|c|}{\textbf{Equilibrium B}} \\ \cline{2-5}
    & $E_\textrm{pot}$ & $E_\textrm{ZPE}^\textrm{DFT} + \int_0^T C_{V,\textrm{vib}}dT$ & $T\times S_\textrm{rot}$ & $T\times S_\textrm{vib}$ \\ \hline
    \textbf{DFT} & \textbf{-56.095} & 0.342 & 0.267 & 0.511 \\ \hline
    \textbf{Hybrid DFT-DMC} & \textbf{-56.093(0)} & (same as DFT) & 0.267(0) & (same as DFT) \\ \hline
    \textbf{CCSD} & \textbf{-55.985} & 0.345 & 0.266 & 0.477 \\ \hline
\end{tabular}
\caption{Breakdown of contributions to free energy for the equilibrium states and transition state of the S$_\textrm{N}$2 system. The dominant contribution (highlighted in bold) to $\Delta G^\oplus$ is the difference in potential energy between the transition state and the equilibrium state. Since this quantity is computed using DMC in the hybrid DFT-QMC approach, the free energy difference between the equilibrium and transition states in the hybrid approach shows greater accuracy than DFT and matches the CCSD result more closely. All energies are in eV and $T=298$ K.}
\label{tab:SN2_thermodynamic_breakdown}
\end{table}